\documentclass[12pt]{article}
\usepackage{amsmath}
\usepackage{amssymb,amsfonts}
\usepackage[all]{xy}
\usepackage{graphicx}
\usepackage[parfill]{parskip}
\usepackage[usenames]{color}
\usepackage{sidecap}
\usepackage{pdfpages}

\numberwithin{equation}{section}
\numberwithin{table}{section}

\newcommand{\be}{\begin{equation}}
\newcommand{\ee}{\end{equation}}
\def\beq{\begin{eqnarray}}
\def\eeq{\end{eqnarray}}
\def\ba{\begin{eqnarray}}
\def\ea{\end{eqnarray}}

\def\ep1{\epsilon_1}
\def\eps2{\epsilon_2}

\newcommand{\IZ}{\mathbb{Z}}

\newcommand{\IN}{\mathbb{N}}
\newcommand{\IR}{\mathbb{R}}

\newcommand{\tr}{\mathrm{Tr\, }}
\newcommand{\im}{\mathrm{Im\, }}
\newcommand{\re}{\mathrm{Re\, }}

\newcommand{\ztop}{\mathrm{Z_{top}}}
\newcommand{\zinst}{\mathrm{Z_{inst}}}

\newcommand{\nn}{\nonumber}

\newcommand{\cN}{{\cal N}}

\newcommand{\cS}{{\cal S}}

\newcommand{\cO}{{\cal O}}

\newcommand{\cF}{{\cal F}}

\newcommand{\aex}{a_{\mbox{\tiny ex}}}
\newcommand{\acft}{a_{\mbox{\tiny cft}}}
\newcommand{\asw}{a_{\mbox{\tiny SW}}}

\newcommand{\ooe}{\frac{1}{\epsilon}}
\newcommand{\thetaf}{\nu}
\newcommand{\phisw}{\phi_{\mbox{\tiny SW}}}
\newcommand{\dualeps}{y}

\begin{document}

\thispagestyle{empty}

\vskip 3cm
\noindent
{\LARGE \bf Pure ${\cal N}=2$ Super Yang-Mills and Exact WKB}

\vskip .1cm
%\begin{center}
\linethickness{.05cm}
\line(1,0){447}
%\end{center}
\vskip .5cm
\noindent
{\large \bf Amir-Kian Kashani-Poor and Jan Troost}
 
\noindent

\vskip 0.15cm
{\em \hskip -.05cm Laboratoire de Physique Th\'eorique\footnote{Unit\'e Mixte du CNRS et
    de l'Ecole Normale Sup\'erieure associ\'ee \`a l'Universit\'e Pierre et
    Marie Curie 6, UMR
    8549.}}
    \vskip -0.22cm

{\em \hskip -.05cm Ecole Normale Sup\'erieure}
   \vskip -0.22cm

{\em \hskip -.05cm 24 rue Lhomond, 75005 Paris, France}

\vskip 1cm

\vskip0cm

\noindent {\sc Abstract:} 
We apply exact WKB methods to the study of the partition function of  pure ${\cal N}=2$ $\epsilon_i$-deformed gauge theory in four dimensions in the context of the 2d/4d correspondence. We study the partition function
at leading order in $\epsilon_2/\epsilon_1$ (i.e. at large central charge) and in an expansion in $\epsilon_1$. We find corrections of the form $\sim \exp[-\frac{\mbox{\tiny SW periods}}{\epsilon_1}]$ to this expansion. We attribute these to the exchange of the order of summation over gauge instanton number and over powers of $\epsilon_1$ when passing from the Nekrasov form of the partition function to the topological string theory inspired form. We conjecture that such corrections should be computable from a worldsheet perspective on the partition function. Our results follow upon the determination of the Stokes graphs associated to the Mathieu equation with complex parameters and the application of exact WKB techniques to compute the Mathieu characteristic exponent.

 \vskip 1cm

\pagebreak

\tableofcontents

\section{Introduction}
Non-perturbative completions of string perturbation theory have been proposed in various backgrounds. One prominent avenue to advance beyond perturbative results is via the holographic correspondence between gauge theories and theories of gravity with a negative cosmological constant. Another approach (with holographic applications) which has proved fruitful is to map
string theories or some of their observables to matrix models and to compute non-perturbative corrections using techniques such as localization.
Nevertheless, the non-perturbative definition of string theory in generic backgrounds
 remains a wide open problem.   
 
Topological string theory is a promising framework within which to tackle this problem, as the nature of the perturbation theory is the same as that of the full string theory -- giving rise to a generically non-convergent genus expansion explicitly linked to computations on underlying Riemann surfaces. Any indication as to what type of approximation to a non-perturbative theory gives rise to such a perturbation theory could help clarify the structures underlying string theory. For a large class of backgrounds, the topological string theory partition function $\ztop$ is fully computable in various series expansions. Lifting these expansions to analytic functions would take us a long way towards a non-perturbative understanding of the theory.

The gauge theory limit of certain topological string theories is $(\epsilon_1,\epsilon_2)$-deformed
 ${\cal N}=2$ supersymmetric Yang-Mills theories in four dimensions. The genus expansion parameter $g_s$ of the topological string is encoded in the $\epsilon_i$-parameters via $g_s^2= \epsilon_1 \epsilon_2$. The problem of determining the instanton partition function $\zinst$ of these theories, which descends from the topological string partition function $\ztop$, has been solved  in \cite{Nekrasov:2002qd}. The solution is presented as a power series in the instanton counting parameter $\Lambda$, with coefficients that are rational functions of $\epsilon_1$ and $\epsilon_2$. The parameter $\Lambda$ reflects the coupling of the gauge theory, which maps to certain K\"ahler parameters in the string theory setting. 
The disadvantage of the resummation of the $\epsilon_i$-series is that it occurs at the expense of introducing this new expansion. For instance, modular properties of the expansion coefficients of the $\epsilon_i$-series are masked in the $\Lambda$-series.

Yang-Mills theories with $\cN=2$ supersymmetries  are related to two-dimensional conformal field theory
 via the two-dimensional / four-dimensional correspondence  \cite{Alday:2009aq}. The powerful computational techniques that exist in the framework of two-dimensional conformal field theory can thus be put to use to elucidate $\epsilon_i$-deformed gauge theories, and by extension, topological string theory.

The conformal field theory technique at the heart of the analysis in this paper relies on null vector decoupling. It permits the computation of conformal blocks, mapped to $\zinst$ under the 2d/4d correspondence, via solution of a differential equation. It has been shown \cite{Mironov:2009uv,KashaniPoor:2012wb,Kashani-Poor:2013oza} that the WKB analysis of this equation, in the limit $\epsilon_2 \rightarrow 0$, reproduces the non-convergent $\epsilon_1$-expansion of $\zinst$. Methods exist to enhance WKB results non-perturbatively (see e.g. \cite{ZinnJustin:2004ib,ZinnJustin:2004cg,Voros,Jidoumou,DDP1,DDP2,DP,KT,IN,Costin}). The point of departure of this paper is to ask what these methods can teach us about the nature of the $\epsilon_i$-expansion of the instanton partition function $\zinst$ and eventually the topological string partition function $\ztop$. Indeed, our analysis will yield corrections to this expansion in the form $\sim e^{-\frac{1}{\epsilon_1}}$.
As the instanton partition function $\zinst$ in the formulation of \cite{Nekrasov:2002qd} is thought to converge \cite{MR2275703}, we conclude that these non-perturbative corrections arise when the order of  summation over powers of the instanton counting parameter $\Lambda$ and over powers of $\epsilon_1$ is inverted. At the locus $\epsilon_1=-\epsilon_2$,
we have a  worldsheet description of the topological string theory, and it  gives rise to a (non-convergent) expansion of $\zinst$ with this reversed order of summation. Once a worldsheet description of the general $\epsilon_i$-deformed theory is formulated (see \cite{Morales:1996bp,Antoniadis:2010iq,Nakayama:2011be,Antoniadis:2013bja,Antoniadis:2013mna} for attempts in this direction), one should seek to compute  the non-perturbative corrections in $\epsilon_1$ within that framework. Here, we determine them by conformal field theory and exact WKB methods. As $\zinst$ in an $\epsilon_i$-expansion remains divergent when restricted to the conventional topological string locus $\epsilon_1=-\epsilon_2$, similar non-perturbative corrections in $\epsilon_i$ should also arise in this more standard setting. 

The theory we shall study is pure ${\cal N}=2$ gauge theory in four dimensions, whose instanton partition function maps to 
an irregular conformal block in two-dimensional conformal field theory \cite{Gaiotto:2009ma}. The relevant null vector decoupling equation maps to the Mathieu equation with complex parameters. We will perform an exact WKB analysis of this equation, determining the Stokes regions for various complex values of the parameters, and incorporating Stokes phenomena in the computation of the characteristic exponent of its solutions. 
This procedure introduces corrections of the order $\exp[-\frac{1}{\epsilon_1}]$ in the relation between the characteristic exponent, linked to the vacuum expectation value $a$ of the scalar field in gauge theory, and a certain complex parameter $u$ of the equation, which is related to the gauge theory partition function via a Matone-style relation. Inverting this relation maps these corrections to non-perturbative corrections to the $\epsilon_1$-expansion of the partition function as a function of the expectation value $a$.

For other studies of non-perturbative effects in topological string theory,
 see e.g. \cite{Hatsuda:2013oxa}, where non-perturbative contributions to the spherical partition function of ABJM theory are mapped to the topological string amplitude on $\mathbb{P}^1 \times \mathbb{P}^1$ together with an additional contribution from the refined topological string amplitude on the same manifold in the Nekrasov-Shatashvili limit \cite{Nekrasov:2009rc}. The authors propose this combination as a non-perturbative definition of the topological string amplitude on this manifold, and conjecture a generalization to arbitrary local Calabi-Yau backgrounds. In \cite{Aniceto:2011nu}, the authors consider the topological string partition function in Gopakumar-Vafa form, closely related to the Nekrasov form of $\zinst$ lifted to topological string theory, in light of resummation techniques. In \cite{Santamaria:2013rua}, the holomorphic anomaly equations are conjectured to hold non-perturbatively, and invoked to conjecture structural properties of a transseries expansion of the topological string free amplitude. These ideas are refined and tested on the example of the topological string on local $\mathbb{CP}^2$ in \cite{Couso-Santamaria:2014iia}, based on computational results for the topological string free energies $F_g$ with $g$ up to $\sim 100$. A similar analysis for the spherical partition function of $\cN=2$ superconformal gauge theory and $\cN=2^*$ gauge theory was performed in \cite{Aniceto:2014hoa}. The paper \cite{Basar:2015xna} studies the Mathieu equation in the context of exact WKB and the 2d/4d correspondence, as we shall do in the following; unlike the present paper, it restricts to $u\in \IR$, thus centering the discussion around the band structure of the equation that is specific to this domain.

The paper is structured as follows. In section \ref{2d4d}, we review the relation between the instanton partition function of $\epsilon_i$-deformed
pure ${\cal N}=2$ super Yang-Mills theory and irregular conformal blocks. We recall the null vector decoupling equation the latter satisfy and map it to a standard form of the Mathieu equation. Section \ref{Mathieu} is dedicated to the study of this equation. Some known facts about the Mathieu equation are summarized in section \ref{M_Floquet}. In section \ref{MathieuWKB}, we study the WKB approximation to the solution of the Mathieu equation. We go beyond perturbation theory in section \ref{beyondperturbationtheory}. We discuss the Stokes graphs associated to the Mathieu equation
with complex parameters and use exact WKB methods to determine properties of analytic solutions to the differential equation. 
 In section \ref{conceptual}, we discuss the non-perturbative corrections to the exact periodicity of the irregular block, and its
 consequences for non-perturbative corrections to the $\epsilon$-expansion of the instanton partition function. We conclude and list interesting open problems in section \ref{conclusions}. Some technical details are relegated to appendix \ref{local_analysis}, while appendices \ref{num_res} and \ref{mama_hm} are dedicated to numerical results.

\section{Pure ${\cal N}=2$ SYM and Conformal Field Theory}
\label{2d4d}
In this section, we review the relation between the gauge theory instanton partition function of pure $\cN=2$ SYM and the irregular conformal block,
as predicted by the 2d/4d correspondence \cite{Alday:2009aq,Gaiotto:2009ma}.

\subsection{The Seiberg-Witten theory of pure $\cN=2$ SYM} \label{sw_parametrization}
The gauge theory we will be concerned with in this paper is pure $\cN=2$ $SU(2)$ super Yang-Mills theory, i.e. the theory of a single vector multiplet with gauge group $SU(2)$. This is the original theory solved by Seiberg and Witten \cite{Seiberg:1994rs} using geometric methods.
The prepotential of the theory is a function of the vacuum expectation value $a$ of the scalar in the vector multiplet.
The coefficients of all instanton contributions 
were determined in \cite{Nekrasov:2002qd} in a form amenable to direct comparison with conformal field theory.

The Seiberg-Witten curve in \cite{Seiberg:1994rs} was given in the form\footnote{A mass scale $\Lambda$ has here been set to 1. It can easily be introduce by dimensional analysis, with $x$ and $u$ carrying mass dimension 2.}
\begin{eqnarray} \label{sw_curve}
y^2 &=& (x^2-1)(x-u) \, ,
\end{eqnarray}
with Seiberg-Witten differential
\be
\lambda  = \frac{\sqrt{x-u}}{\sqrt{x^2-1}} \,dx \, .
\ee
We can choose branch cuts from $x=-1$ to $x=1$ on the real axis, and from $x=u$ to infinity, not crossing the real axis. A basis of cycles on the Riemann surface (\ref{sw_curve}) is then given by a curve circling the line connecting $x=-1$ and $x=1$ in positive orientation without intersecting the second branch cut, and the curve circling the points $x=1$ and $x=u$, intersecting both branch cuts once and intersecting the first curve with intersection number 1. These choices define the $A$- and the $B$-cycle of the torus respectively. The corresponding Seiberg-Witten periods are
\be
a^{(0)} = \oint_A  \lambda \,, \quad a_D^{(0)} = \oint_B  \lambda \, .
\label{periodintegrals}
\ee
The superscripts indicate that these periods constitute the leading terms in formal power series to be introduced shortly.

\begin{SCfigure} \label{q_plane}
\includegraphics[scale=.6]{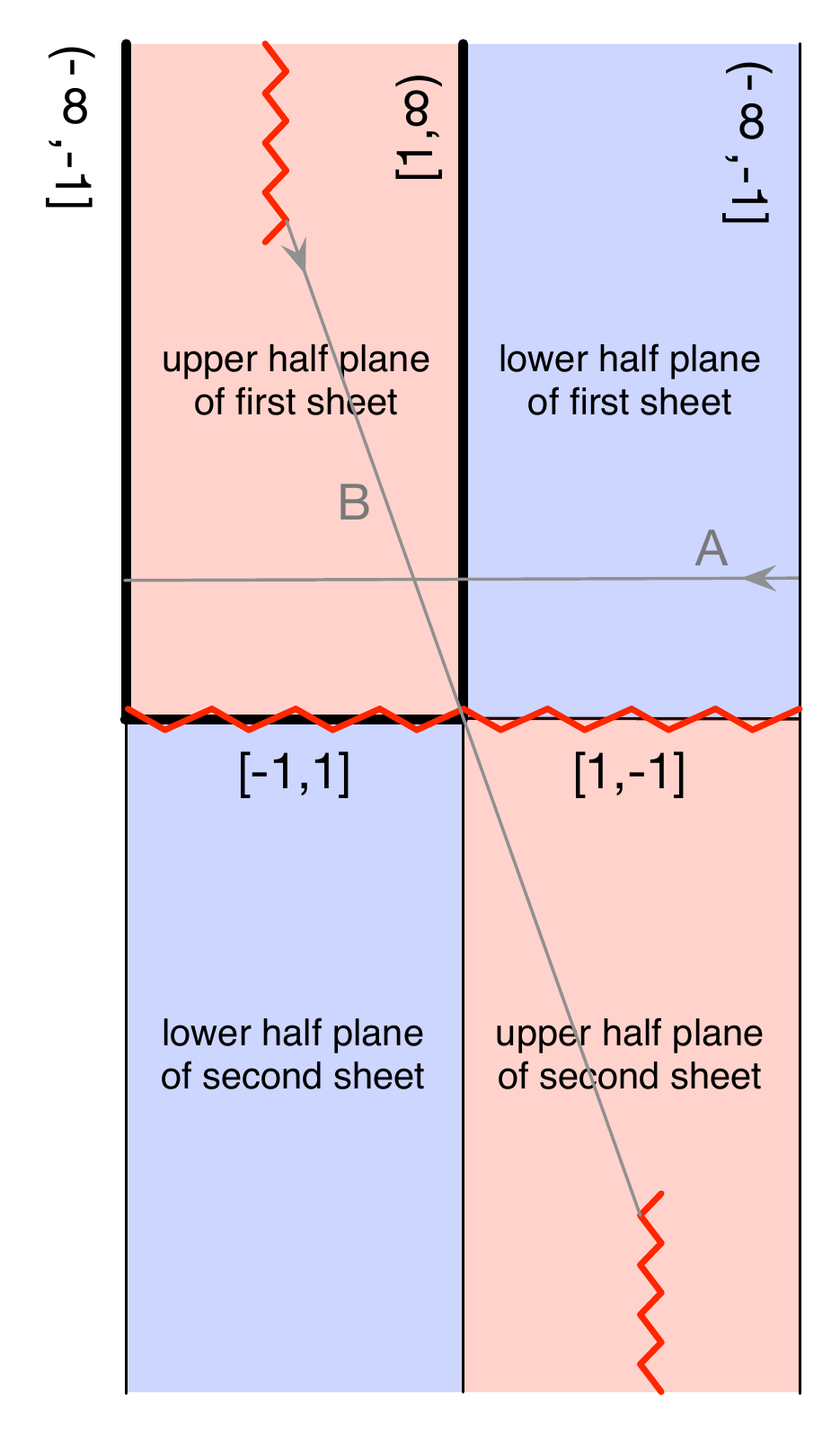}
\centering
\caption{The q-plane. The solid black line is mapped to the real axis of the $x$-plane. The interval $[-\pi, \pi]$ is the double cover of the branch cut $[-1,1]$ in the $x$-plane.} 
\end{SCfigure}

Based in part on previous work \cite{Witten:1997sc,Gaiotto:2009hg}, Gaiotto in \cite{Gaiotto:2009we} suggested shifting the emphasis onto quadratic differentials when analyzing $\cN=2$ theories. In the case at hand,
the quadratic differential reads
\be
\phisw = \frac{x-u}{x^2-1} \, dx \otimes dx \,.
\ee
For our upcoming analysis, the variable redefinition
\be
x = \cos q
\ee 
will prove useful, with regard to which $\phisw$ takes the form
\be \label{phi_sw}
\phisw = (u-\cos q ) \, dq \otimes dq \,.
\ee
The relation between the variables $x$ and $q$ is one-to-one if we restrict the variable $q$ to 
the range $\re q \in [-\pi, \pi
]$, $\im q \ge 0$ -- we will refer to this region as the fundamental domain of the $q$-plane -- and identify the half-lines $\re q = - \pi$ and $\re q = \pi$, as well as the intervals $[-\pi,0]$ and $[0,\pi]$ via the map $q \mapsto -q$. 
See figure \ref{q_plane}.
The branch cut connecting the points $x = \pm 1$ maps to the intervals $[-\pi,0]$ and $[0,\pi]$ of the $q$-plane, undoing the identification of these two intervals (as the factor $\sqrt{1-x^2} = \sin q$ in the curve variable $y$ differs by a sign between previously paired points). The second branch cut runs from the preimage $q_u$ of $u$ in the fundamental domain to positive imaginary infinity, without crossing the imaginary axis. We can identify the second sheet of the $x$-plane with the image of the fundamental domain of the $q$-plane under the map $q \mapsto -q$. Overall, the Riemann surface (\ref{sw_curve}) in the $q$ coordinate is hence given by a cylinder with branch cuts emanating from $q_u$ and $-q_u$ towards the two ends of the cylinder. In this representation of the Riemann surface, the $A$-cycle on the $q$-plane is represented by a curve running from $ \pi + i |q_0|$ to $-\pi + i |q_0|$ without intersecting the second branch cut, or simply by a cycle of the cylinder. The $B$-cycle is represented by a curve running from a point on the second branch cut in the fundamental domain of the $q$-plane to its image point in the lower half-plane, crossing both branch cuts once. 

A third choice of variable, underlying the analysis in \cite{Gaiotto:2009ma,Gaiotto:2012sf} relating pure $\cN=2$ $SU(2)$ gauge theory to conformal field theory, is given by
\be
z = e^{iq} \,.
\ee
In terms of $z$, the quadratic differential is (up to an overall factor, and with the scale $\Lambda$ re-introduced, as it will come in handy in subsection \ref{irreg_block}) given by
\be \label{qd_in_z}
\phisw = \left( \frac{\Lambda^2}{z^3} - \frac{2u}{z^2} + \frac{\Lambda^2}{z} \right) \, dz \otimes dz \,.
\ee

\subsection{The irregular conformal block} \label{irreg_block}
The original 2d/4d correspondence \cite{Alday:2009aq} conjectured a relation between the instanton partition function of superconformal $\cN=2$ gauge theories and Virasoro conformal blocks. This relation was extended to asymptotically free gauge theories in \cite{Gaiotto:2009ma} by introducing irregular conformal blocks. These are defined as the norm of so-called Gaiotto vectors: formal power series $|\Delta, \Lambda^2 \rangle$ in the parameter $\Lambda^2$ with coefficients valued in the Verma module of highest weight $\Delta$ satisfying the relations \cite{Gaiotto:2009ma}
\be
L_1 |\Delta, \Lambda^2 \rangle  = \Lambda^2  |\Delta, \Lambda^2 \rangle  \,, \quad L_2 |\Delta, \Lambda^2 \rangle = 0 
\ee
order by order in $\Lambda^2$. The requirements on the states $|\Delta, \Lambda^2 \rangle$ are chosen such that the ensuing expectation value of the energy momentum tensor in the semi-classical limit reproduces the quadratic differential (\ref{qd_in_z}), following the prescription of \cite{Alday:2009aq}, further elucidated in \cite{Kashani-Poor:2014mua}.

Gaiotto states can be constructed via collision of primary fields \cite{Gaiotto:2012sf}.
In this manner, the irregular conformal block for pure $SU(2)$ can be obtained by a limiting procedure from the conformal block on the sphere with four 
insertions \cite{Gaiotto:2012sf}. Starting from the conformal Ward identity satisfied by the product of two primaries 
$\Psi_{\Delta_i}$ acting on the vacuum $|0 \rangle $,
\begin{eqnarray}
T_{>} (z) \Psi_{\Delta_2} (z_2) \Psi_{\Delta_1} (z_1) | 0 \rangle
&=& (\frac{\Delta_2}{(z-z_2)^2}  + \frac{1}{z-z_2} \partial_2 + \frac{\Delta_1}{(z-z_1)^2}  + \frac{1}{z-z_1} \partial_1 )
\times
\nonumber \\
& & 
\Psi_{\Delta_2} (z_2) \Psi_{\Delta_1} (z_1) | 0 \rangle \, ,
\end{eqnarray}
where $T_{>}$ indicates the sum over modes of the energy-momentum tensor $T$ with weight larger or equal than $-1$,
we can send the insertion $z_1 \rightarrow 0$ to generate a primary state $| \Delta_1 \rangle$. We moreover use the fact that $L_{-1}=\partial_1+\partial_2$ generates a translation on the correlator, as well as the parameterizations
\begin{eqnarray}
Q&=& b + b^{-1} \nonumber \\
\Delta_i &=& \alpha_i( Q- \alpha_i)
\nonumber \\
c_1 &=& - z_2 \alpha_1
\nonumber \\
\alpha &=& \alpha_1 + \alpha_2 \, ,
\end{eqnarray}
to find, in the second limit $z_2 \rightarrow 0$ and $\alpha$ and $c_1$ fixed,
\begin{eqnarray}
T_{>} (z) \Psi_{\Delta_2} (z_2) \Psi_{\Delta_i} (z_1) | 0 \rangle
&=& (-\frac{c_1^2}{z^4} + \frac{2 c_1(Q-\alpha)}{z^3} + \frac{ \alpha (Q-\alpha)}{z^2}
+ \frac{1}{z^2} c_1 \partial_{c_1} + \frac{1}{z} L_{-1} ) \times
\nonumber \\
& & 
\lim_{z_2 \rightarrow 0} ( z_2^{2 \alpha_1 \alpha_2} 
\Psi_{\Delta_2} (z_2) | \Delta_1 \rangle ) \, .
\end{eqnarray}
To relate this expression to the quadratic differential (\ref{qd_in_z}) with third order pole, we  take the further limit $\Lambda_2=-c_1^2 \rightarrow 0$
while $\Lambda_1 = 2 c_1 (Q-\alpha)$ is kept finite. The state that this limiting procedure gives rise to satisfies the Whittaker properties
\be
L_n | \Delta , \{\Lambda_i\} \rangle = \Lambda_n | \Delta , \{\Lambda_i\} \rangle
\ee
with $\Lambda_1=\Lambda^2$ and $\Lambda_n=0$ for $n>1$ \cite{Gaiotto:2009ma}. 

By explicit calculation to a fixed order, it was demonstrated  \cite{Gaiotto:2009ma}  that the norm of this state coincides with the instanton partition function of $\epsilon_i$-deformed pure ${\cal N}=2$ super Yang-Mills theory
\begin{equation} \label{2d4d_id}
\langle \Delta, \Lambda | \Delta, \Lambda \rangle = \zinst^{ {\cal N} =2, SU(2) \mbox{\tiny pure}}(\acft)
\end{equation}
with the conventional 2d/4d identifications
\be
b^2 = \frac{\epsilon_2}{\epsilon_1} \,, \quad p = \frac{\acft}{2\pi i \sqrt{\epsilon_1 \epsilon_2}} \,,
\ee
where the momentum $p$ is related to the conformal weight $\Delta$ by the formula $\Delta (ip) = \frac{Q^2}{4} + p^2$. We will introduce numerous variables closely related to the vacuum expectation value of the adjoint
scalar in the gauge theory, and have therefore denoted the conformal field theory quantity determining the weight of the exchanged state with the subscript
$\mbox{`cft'}$.

\subsection{The null vector decoupling equation}
Our point of departure will be the fact that the irregular block after insertion of a degenerate operator satisfies a null vector decoupling equation. 
We denote the irregular block with insertion $\Phi_{h_{2,1}}$ at the point $z$ as $\Psi(z)$,
\be
\Psi(z) =  \langle  \Delta (\frac{\acft}{2 \pi i \sqrt{\epsilon_1 \epsilon_2}}-\frac{b}{4}) , \Lambda | \Phi_{h_{2,1}} (z) | \Delta ( \frac{\acft}{ 2 \pi i  \sqrt{\epsilon_1 \epsilon_2}}+\frac{b}{4}) , \Lambda \rangle \,.  \label{deg_block}
\ee
In this definition, we slightly shift the momenta of the incoming and outgoing irregular vectors away from $\frac{\acft}{2 \pi i \sqrt{\epsilon_1 \epsilon_2}}$ such that
\be
\Delta(\frac{\acft}{2 \pi i  \sqrt{\epsilon_1 \epsilon_2}}+\frac{b}{4}) -\Delta(\frac{\acft}{ 2 \pi i  \sqrt{\epsilon_1 \epsilon_2}}-\frac{b}{4})= \frac{\acft}{2 \pi i  \epsilon_1} \,.
\ee
This condition is necessary for the correlator not to vanish.
The derivation of the null vector decoupling equation this correlator satisfies was worked out in e.g.  \cite{Awata:2009ur,Awata:2010bz}. After a change of
variables $z = e^w$ and the rescaling
\be
\Psi(z) = z^{\Delta (\frac{\acft}{2 \pi i  \sqrt{\epsilon_1 \epsilon_2}}-\frac{b}{4})-\Delta ( \frac{\acft}{2 \pi i  \sqrt{\epsilon_1 \epsilon_2}}+\frac{b}{4})-h_{2,1}} Y(w)  \, ,
\label{redef1}
\ee
the equation reads
\be
(\frac{\epsilon_1}{\epsilon_2} \partial_w^2 + \frac{\acft}{\pi i \epsilon_2} \partial_w + \frac{ \Lambda^2}{\epsilon_1 \epsilon_2} (e^w + e^{-w})
+ \frac{\Lambda}{4} \partial_\Lambda ) Y (w)
=  0 \, .
\ee
To simplify the equation further, we define
\be
\Xi(w) = e^{ \frac{2 \acft^2}{\pi i \epsilon_1 \epsilon_2} \log \Lambda} e^{- \frac{\acft}{2 \pi i \epsilon_1} w} Y(w) \, . \label{redef2}
\ee
As the monodromy of $\Psi(z)$ that follows from its definition (\ref{deg_block}) is accounted for by the prefactor in (\ref{redef1}), the function $\Xi(w)$ has periodicity $\Xi(w - 2\pi i) = e^{\frac{\acft}{\epsilon_1}} \Xi(w)$. It satisfies the  equation
\be
(\frac{\epsilon_1}{\epsilon_2} \partial_w^2  + \frac{\Lambda^2}{\epsilon_1 \epsilon_2} (e^w + e^{-w})
+ \frac{\Lambda}{4} \partial_\Lambda ) \,\Xi(w) = 0 \, . 
\label{Toda}
\ee
This differential equation for $\Xi(w)$ is  exact,  both in $\epsilon_1$ and in $\epsilon_2$.

\subsection{The semi-classical limit in the central charge}
To be able to extract the conformal block (\ref{2d4d_id}) of interest from solutions of the null vector decoupling equation (\ref{Toda}), we will study it
to leading order in $\epsilon_2/\epsilon_1$. This is a first WKB approximation. The limit renders the Gaiotto state heavy compared
to the light degenerate insertion, justifying the factorization ansatz
\be
\Xi(w)= e^{ \frac{1}{ \epsilon_1 \epsilon_2} {\cal F} (\Lambda)} \chi(w|\Lambda) \,,
\ee
with $\log \chi(w|\Lambda)$ finite in the limit $\epsilon_2 \rightarrow 0$. The first factor is to be identified to the irregular block without the degenerate insertion, while the second factor is associated
to the light degenerate mode. Plugging the ansatz into  equation (\ref{Toda}), we obtain
\be
(\frac{\epsilon_1}{\epsilon_2} \partial_w^2  + \frac{\Lambda^2}{\epsilon_1 \epsilon_2} (e^w + e^{-w})
+ \frac{\Lambda}{4} \partial_\Lambda + \frac{1}{4 \epsilon_1 \epsilon_2} \partial_{\log \Lambda} {\cal F}) \chi(w|\Lambda) = 0 \, .
\ee
We  multiply the equation by $\epsilon_1 \epsilon_2/
\Lambda^2$ to find
\be
(\frac{\epsilon_1^2}{\Lambda^2} \partial_w^2  +  (e^w + e^{-w})
 + \frac{1}{4 \Lambda^2 } \partial_{\log \Lambda} {\cal F}) \chi(w|\Lambda) = 0 \, .
 \label{Toda2}
\ee
The differential equation is exact in $\epsilon_1$. It was already studied (pre-2d/4d correspondence!) in \cite{MR2214246} to obtain the prepotential of pure $\cN=2$ $SU(2)$ gauge theory. Upon the variable redefinition
\be
w= i q \, ,  \qquad
\frac{\epsilon_1^2}{2 \Lambda^2} = \epsilon^2 \, , 
\qquad
- \frac{1}{8 \Lambda^2} \partial_{\log \Lambda} {\cal F} = u \, ,
\label{identificationofu}
\ee
it is mapped to the form
\be \label{math_eq}
(- \epsilon^2 \partial_q^2 + \cos q ) \psi(q) = u \, \psi(q) \,.
\ee
This is the Mathieu equation, which has been studied from various angles in the mathematics literature 
(see e.g. \cite{MS,OLBC10}). 

\section{The Mathieu Equation and Exact WKB}
\label{Mathieu}
The technical core of this paper is the exact WKB analysis of the Mathieu equation.  This method will permit us to compute corrections to the formal WKB solutions to the equation, in a sense which we shall make precise in this section. Recall that the formal solutions reproduce the non-convergent $\epsilon$-expansion of the instanton partition function of pure $\cN=2$ gauge theory at large $a/\epsilon$. The correction terms computed here yield information beyond this asymptotic expansion.

 Before turning to the WKB analysis in subsection \ref{MathieuWKB}, we  collect some general results regarding the Mathieu equation and its solutions in the following subsection, based on the exposition in \cite{MS}.

\subsection{The Mathieu equation} \label{M_Floquet}
The Mathieu equation (\ref{math_eq}) is an ordinary differential equation of degree 2. It has a unique solution upon specifying two boundary conditions on the solution $\psi$, e.g. the values of $\psi$ and its derivative $\psi'$ at a point. The solution is entire in the variable $q$ as well as the parameters $u \epsilon^{-2}$
and $\epsilon^{-2}$ \cite{MS}. As the potential term in the Mathieu equation is periodic (of period $2 \pi$ in the variable $q$), Floquet theory applies. In particular, two independent solutions of the equation exist that satisfy
\begin{equation}   \label{Floquet}
\left( 
\begin{array}{c}
\psi \\
\psi'
\end{array}
\right)(q) = e^{i \thetaf q/2} { \chi}(q) \,,
\end{equation}
for appropriate $\thetaf$, with ${\chi}$ a $2\pi$ periodic vector valued function of $q$. Such solutions are called Floquet solutions. The factor
$e^{i \thetaf \pi}$, which reflects the monodromy of the solution $\psi$ under $q \rightarrow q + 2\pi$,  is called a characteristic multiplier of the equation, and $\thetaf$ is a characteristic exponent. Since
the characteristic exponent is defined via the $\log$ of the characteristic multiplier, it is only defined modulo $2 \mathbb{Z}$. In  equation (\ref{Floquet}), shifting the characteristic exponent $\thetaf$ by $2 \mathbb{Z}$ requires rescaling the periodic function ${\chi}(q)$. We introduce the parameter $\hat{\nu}$ by writing $\thetaf = \hat{\nu} + 2n$ for $n \in \mathbb{Z}$ and $\re(\hat{\nu}) \in [-1,1]$. The characteristic exponents $\hat{\nu}$ of the two independent Floquet solutions of the Mathieu equation add to $0$, by the absence of a first order derivative term in (\ref{math_eq}). We can express the Floquet solutions as linear combinations $\psi = A \psi_1 + B \psi_2$ of two solutions $\psi_1, \psi_2$ satisfying the  boundary conditions
\begin{equation}
\left( 
\begin{array}{cc}
\psi_1 & \psi_2 \\
\psi_1' & \psi_2'
\end{array}
\right)(0) = 
\left( 
\begin{array}{cc}
1 & 0 \\
0 & 1
\end{array}
\right) \,.
\end{equation}
Requiring that coefficients $A$ and $B$ exist such that the corresponding linear combination satisfies the Floquet monodromy condition with characteristic exponent $\thetaf$ entails the following constraint on the characteristic exponent: 
\begin{equation} \label{ce}
\cos \pi \thetaf = \psi_1(u,\epsilon,\frac{\pi}{2})  \,.
\end{equation}
This equation is called the characteristic equation of the Mathieu equation, or somewhat less fortuitously, a quantization condition. For fixed parameters $u$ and $\epsilon$, equation (\ref{ce}) determines the characteristic exponent up to a sign and $2 \mathbb{Z}$ ambiguity. As we have seen, the $2 \mathbb{Z}$ ambiguity corresponds to a rescaling of the periodic function $\chi(q)$ in (\ref{Floquet}) (or a relabeling of the Fourier coefficients of $\psi$), whereas the two signs correspond to the two independent Floquet solutions. 

In applications, it is often important to know the possible values of the parameter $u$ that are consistent with given
values of the parameter  $\epsilon$ and the characteristic exponent $\thetaf$. When the Mathieu equation arises as a 
Schr\"odinger equation in quantum mechanics, fixing $\thetaf$ corresponds to fixing the periodicity condition on the wave function, and determining the values of $u$ that permit solutions with this periodicity is tantamount to determining the spectrum of the Hamiltonian. In the context of this paper, $\thetaf$ corresponds to the exchanged momentum of the conformal block $\Xi(w)$ in (\ref{redef2}) via
\be
i \thetaf \pi = \frac{\acft}{\epsilon_1} \,.
\ee
By the 2d/4d correspondence, the characteristic exponent hence maps to the scalar vacuum expectation value of the adjoint scalar in the vector multiplet of the $SU(2)$ gauge theory. The corresponding $u$ determines the partition function of the gauge theory via the generalized Matone relation \cite{Matone:1995rx,Flume:2004rp,Kashani-Poor:2014mua} as it arises in (\ref{identificationofu}), $ u = -\frac{1}{8\Lambda^2} \Lambda \frac{ \partial \cF}{\partial \Lambda}$. We will review in subsection \ref{MathieuWKB} that the characteristic exponent at fixed $u$ is approximated by the $A$-period of the Seiberg-Witten differential in a WKB analysis of the Mathieu equation, and study how exact WKB methods permit determining corrections to this relation in section \ref{bpt}. In the rest of this subsection, we will review what can be learned about the relation between $\thetaf$ and $u$ from the study of the characteristic equation (\ref{ce}), following the classic reference \cite{MS}.

For a given $\epsilon$ and a  non-integer $\thetaf$, a discrete infinite number of solutions of the equation (\ref{ce}) for $u$ exists. We can label these as $u(\thetaf, N, \epsilon)$, with $N \in \IZ$. Notice that by the periodicity of the cosine function, we can define an integer-valued function $N(M_1,M_2)$ on $\IZ \times \IZ$ such that $u(\thetaf+ 2M_1, M_2, \epsilon)= u(\thetaf, N(M_1,M_2),\epsilon)$ for $M_1,M_2 \in \IZ$. One can moreover show that the choice $N(M_1,M_2) =2(M_1+M_2)$ is possible, allowing us
 to combine the variables $\thetaf$ and $N$ and express the solution $u$ as a function $u(\thetaf+2N,\epsilon)$. No generality is lost by calling the first variable $\thetaf$. Recall that by Floquet theory, 
 we furthermore have the parity property $u(\thetaf,\epsilon) = u(-\thetaf, \epsilon)$. 

\begin{figure}
\includegraphics[scale=.5]{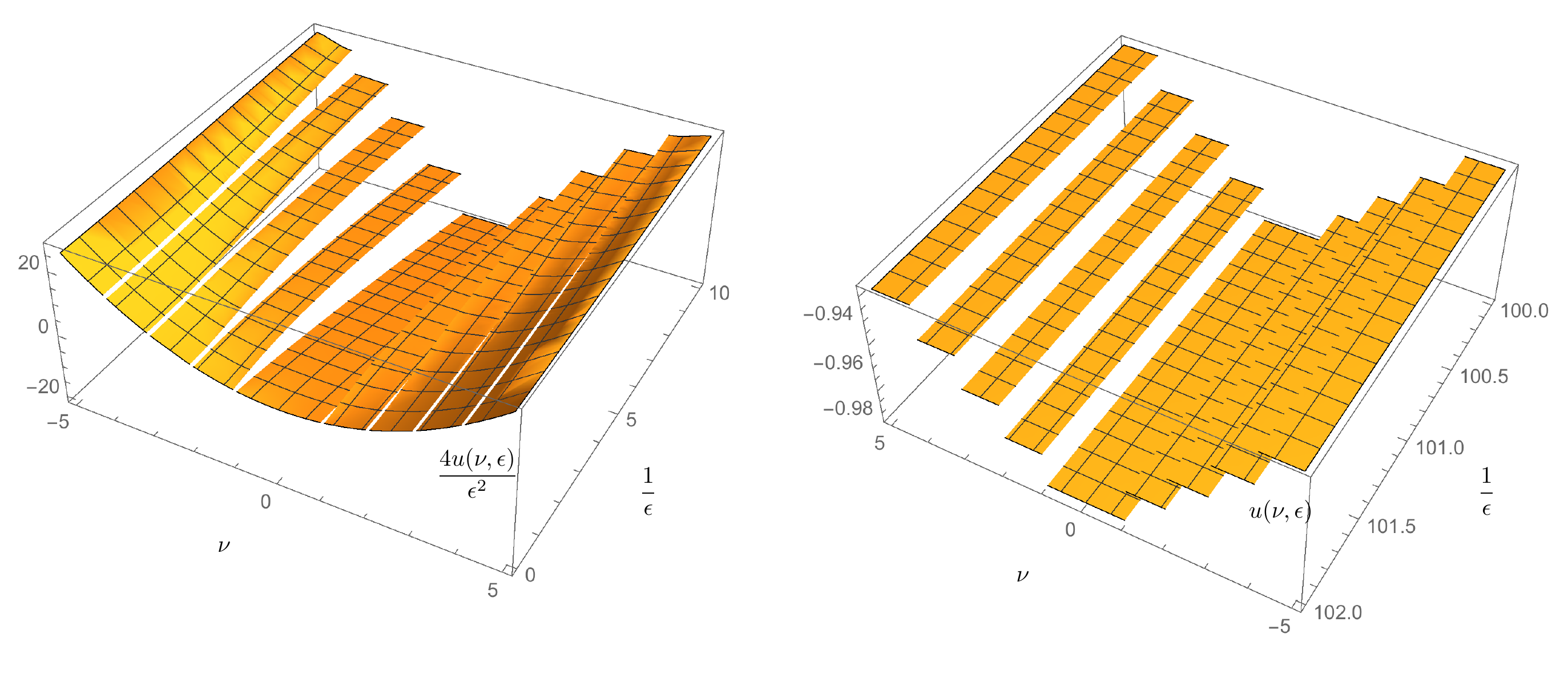}
\centering
\caption{The variable $u$ is discontinuous as a function of real $\thetaf$. On the left, the behavior $\frac{4 u}{\epsilon^2} \sim \nu^2$ is visible, as well as the narrowing of the bands as $\epsilon \rightarrow \infty$. On the right, the limit $u \rightarrow -1$ as $\epsilon \rightarrow 0$ is apparent.} \label{bands}
\end{figure}

At integer $\thetaf=n \in \IZ - \{0\}$, two discrete infinite families of solutions to the characteristic equation (\ref{ce}) exist, labelled as $u_+(n,\epsilon)$ and $u_-(n,\epsilon)$  (these solutions, rescaled by the factor $4 \epsilon^{-2}$, are usually denoted $a_n$ and $b_n$). The function $u(\thetaf,\epsilon)$ for $\thetaf \in \IR$ is discontinuous at $\thetaf \in \IZ$, 
\be
\lim_{\thetaf \rightarrow n^\pm} u(\thetaf,\epsilon) = u_\pm(n,\epsilon)  \,,
\ee
with ${\thetaf \rightarrow n^\pm}$ indicating that $n \in \IZ$ is approached from above/below respectively. These  discontinuities along the real axis give rise to what is referred to as the band structure of the spectrum of the Mathieu equation, see figure \ref{bands}.

Convergent and asymptotic series in $\epsilon$ for $u$ are known, at large and small $\epsilon$ respectively. It can be uniquely characterized by its convergent power series expansion around $\epsilon = \infty$,
\be \label{series_non_int}
\frac{1}{\epsilon^2}u(\thetaf, \epsilon) = \frac{\thetaf^2}{4} + \frac{1}{4(\thetaf^2-1)}\frac{1}{\epsilon^2}+\cO(\frac{1}{\epsilon^4}) \,.
\ee
The coefficients of the series expansion can be found by equating negative powers of $\epsilon$ in the following continued-fraction equation. With the notation $x = 4 u \epsilon^{-2}$,
\be \label{cont_frac_non_int}
x- \thetaf^2 - \frac{\epsilon^{-2}}{x-(\thetaf+2)^2-}\frac{\epsilon^{-2}}{x-(\thetaf+4)^2-}\cdots = \frac{\epsilon^{-2}}{x-(\thetaf-2)^2-}\frac{\epsilon^{-2}}{x-(\thetaf-4)^2-}\cdots \,.
\ee
The series that results from this procedure has finite convergence radius when the parameter $\thetaf$ is not
an integer. For integer $\thetaf$, convergent power series expansions around $\epsilon = \infty$ exist as well, with leading term
\be \label{series_int}
\frac{1}{\epsilon^2}u_{\pm}(n, \epsilon) = \frac{n^2}{4} + \cO(\frac{1}{\epsilon^2}) \,,
\ee
but the corresponding coefficients are given by recourse to different continued-fraction equations. E.g., for $u_+$ and even $n$, 
\begin{multline}  \label{cont_frac_int}
x -(2n)^2 - \frac{\epsilon^{-2}}{x-(2n-2)^2-}\frac{\epsilon^{-2}}{x-(2n-4)^2-}\cdots \frac{\epsilon^{-2}}{x-2^2-}\frac{2 \epsilon^{-2}}{x} = \\
 \frac{\epsilon^{-2}}{(2n+2)^2-x-}\frac{\epsilon^{-2}}{(2n+4)^2-x-} \cdots \,.
\end{multline}
Notice that this continued-fraction equation is different in nature from the one appearing in equation (\ref{cont_frac_non_int}), as the variable $n$  also appears as a summation index. Hence, the equation yields the coefficients of the series expansion only once $n$ has been assigned a value. For large $n$, the leading coefficients in the series expansion can nevertheless be expressed as rational functions of the parameter $n$, and they coincide with the coefficients for $\thetaf$ non-integer, evaluated at $\thetaf=n$. This agreement breaks down for higher terms in the series, rendering (\ref{series_int}) convergent, whereas (\ref{series_non_int}) diverges for integer $\thetaf$.

For $\epsilon^2$ real and small, it is known that the values of the two solutions
$u_+(n,\epsilon)$ and $u_-(n+1,\epsilon)$ approach each other and have the asymptotic expansion, for $\epsilon^2$ positive,
\be \label{asym_small_eps}
u_+(n,\epsilon) \sim u_-(n+1,\epsilon) \sim -1 + \frac{s}{\sqrt{2}}\epsilon - \frac{1}{32 }(s^2+1) \epsilon^2 + \ldots \,,
\ee
where $s= 2 n+1$, $n=0,1,2, \ldots$. A similar expansion can be derived for $\epsilon^2$ small and negative, with leading term $+1$. One can show that the difference between $u_+(n,\epsilon)$ and $u_-(n+1,\epsilon)$ scales like $\exp \left( -\frac{1}{|\epsilon|} \right)$. The asymptotic expansion (\ref{asym_small_eps}) is hence also valid for $u(\thetaf,\epsilon)$, with $\thetaf$ real and in the interval $n < \thetaf <n+1$.

Finally, one can ask about the analytic properties of $u(\thetaf,\epsilon)$ as a function of $\thetaf$. According to \cite{MS}, the solution $\psi_1$ on the right hand side of the characteristic equation (\ref{ce}) is analytic as a function of $\thetaf$ and $\epsilon$ away from possible branch cuts. By the implicit function theorem, the parameter $u$ will be analytic as a function of $\thetaf$ and $\epsilon$ away from these branch cuts and from zeros of the $u$-derivative of $F(u,\nu) = \psi_1(u,\epsilon,\frac{\pi}{2}) - \cos \pi \nu$. By writing
\be 
\frac{\partial \psi_1(u,\epsilon,\frac{\pi}{2})}{\partial u} = - \pi \frac{\partial \thetaf}{\partial u} \sin \pi \thetaf \,,
\ee
the discontinuities of $u$ for real $\thetaf$ become visible as the zeros of $\sin \pi \thetaf$. Additional analytic complications will arise at zeros of the factor $\partial_u \thetaf$. Using Mathematica, we found some
evidence for the existence of one stationary point $u_0(n)$ of $\thetaf(u)$ for each pair $u_\pm(n)$, $n \in \IZ$, with $u_-(n) < u_0(n) < u_+(n)$.

\subsection{WKB analysis of the Mathieu equation}
\label{MathieuWKB}
In this section, we will review the WKB approximation to the Mathieu equation. This is standard material. For future convenience, we use the conventions in \cite{IN}. For further results on the WKB analysis of this equation, see \cite{He:2010xa,He:2010if}.

The starting point of the general theory is a second order differential equation of the form
\be
[ \epsilon^2 \partial_q^2  - Q(q,\epsilon) ] \Psi(q)= 0 \,,   \label{gen_diff}
\ee
on a Riemann surface $\Sigma$, depending on a potential $Q(q,\epsilon) = \sum_{k=0}^N Q_k(q) \epsilon^k$ which is a polynomial in $\epsilon$ with coefficients $Q_k(q)$ that are meromorphic on $\Sigma$, satisfying conditions outlined in \cite{IN}. We will immediately specialize to the Mathieu equation in the form
\be
[ \epsilon^2 \partial_q^2  - (\cos q  - u)  ] \Psi(q)= 0 \,,   \label{math_diff}
\ee
with $\Sigma$ chosen as the cylinder $-\pi \le \im q < \pi$ compactified to a sphere by adding points at $\pm i \infty$. The parameters $\epsilon$ and $u$ can be complex. Hence,
\be \label{q0}
Q_0 = \cos q - u
\ee
and, since we pick the parameter $u$ to be $\epsilon$-independent, we have $Q_n=0$ for $n>0$.

The WKB ansatz for the solution of the differential equation is 
\be \label{WKB_general_form}
\psi(q,\epsilon) = \exp ( \int^q S \,dq )  \,,
\ee
with $S$ expanded as a formal power series in $\epsilon$,
\be
S = \frac{1}{\epsilon} S_{-1} + S_0 + \epsilon S_1 + \ldots \,.
\ee
By plugging this ansatz into the differential equation, one immediately derives the following recursion relation for the coefficients $S_{n}$:
\begin{eqnarray}
S_{-1}^2 &=& Q_0 \,, \label{initial}\\
2 S_{-1} S_{n+1} + \sum_{\substack{n_1+n_2=n \\ 0 \le n_j \le n}} S_{n_1} S_{n_2} + \frac{d S_n}{dq} &=& Q_{n+2} \quad \,,\quad n>-1 \,.
\end{eqnarray}
Equation (\ref{initial}) has two solutions, $S_{-1} = \pm \sqrt{Q_0}$. We note that up to normalization, the Seiberg-Witten differential of pure $\cN=2$ $SU(2)$ gauge theory thus arises in the WKB analysis of the Mathieu equation as $\lambda = S_{-1} dq$. In this matching, the variable $u$ defined via (\ref{identificationofu}) coincides with the conventional variable parametrizing the Seiberg-Witten $u$-plane introduced in subsection \ref{sw_parametrization}, a manifestation of a generalized Matone relation.

Depending on the choice of the sign of $S_{-1}$, we obtain two solutions to the recursion relations, which we label $S^\pm$. The first few series coefficients, for $u$ chosen to be $\epsilon$-independent, are given by 
\begin{eqnarray}
S_{-1}^\pm &=& \pm \sqrt{ \cos q -u}  \,,
\nonumber \\
S_0^\pm &=& - \frac{1}{2} d \log S_{-1} / dq  = \frac{1}{4} \frac{\sin q}{ { \cos q -u}}  \,,
\nonumber \\
S_{1}^\pm &=& \pm \frac{ \cos 2 q + 8 u \cos q -9}{64 (\cos q - u)^{5/2}}  \,, 
\nonumber \\
S_{2}^\pm &=&-\frac{\sin (q) \left(20 u \cos (q)+\cos (2 q)+8 u^2-29\right)}{128  (u-\cos (q))^4}  \,,
\nonumber \\
S_{3}^\pm &=& \mp \frac{1}{{16384 (\cos (q)-u)^{11/2}}} \Big(16 \left(32 u^2-265\right) u \cos q+20 \left(112 u^2-173\right) \cos 2 q+ \nonumber\\
&&912 u \cos 3 q+25 \cos 4 q-1344 u^2+5355 \Big) \, .
\nonumber
\end{eqnarray}
Introducing
\be
S_{odd} = \frac{1}{2} (S^+ - S^-) \,,\quad
S_{even} = \frac{1}{2} (S^++S^-) \,,
\ee
one can show that
\be
S_{even} = - \frac{1}{2} \frac{ d \log S_{odd}}{d q}  \,.
\ee
The formal WKB solution (\ref{WKB_general_form}) thus takes the form
\be \label{sqrt_s_odd}
\psi_{\pm}(q,\epsilon) = \frac{1}{\sqrt{S_{odd}(q,\epsilon)}} \exp \left(\pm \int^q S_{odd}(q,\epsilon) \, dq \right)  \,,
\ee
where this formal expression is to be interpreted as an analytic function in $q$ with branch cuts multiplying a formal power series as follows:
\be \label{formal_WKB}
\psi_{\pm}(q,\epsilon) = \exp \left(\pm \frac{1}{\epsilon} \int^q \sqrt{Q_0(q)} \,dq \right) \epsilon^{1/2} \sum_{k=0}^\infty \epsilon^k \psi_{\pm,k}(q) \,.
\ee
To fix the normalization of (\ref{sqrt_s_odd}), we need to specify the starting point of the integral. In this paper, we will choose this starting point to coincide with the zeros of $Q_0$, called turning points. Care is required in defining the ensuing integral \cite{IN}, as the coefficients $S_{n}$, $n\ge0$, have poles at the turning points.

Our analysis will require comparing WKB solutions normalized with regard to different turning points. These are related by exponentials of periods of $S_{odd}$, for which we introduce the notation
\be \label{WKB_periods}
a = \epsilon \int_A S_{odd} \,dq \,, \quad a_D = \epsilon \int_B S_{odd}\,dq \,.
\ee
The integrals are to be understood order by order in $\epsilon$. To leading order, the integrals $a$ and $a_D$ equal the Seiberg-Witten periods $a^{(0)}$ and $a_D^{(0)}$, which can be expressed in terms of hypergeometric functions,
\begin{eqnarray}
a^{(0)} &=& - 2 \pi  i  \sqrt{u+1} \, {}_2F_1\left(-\frac{1}{2},\frac{1}{2};1;\frac{2}{u+1}\right) \,,\\
 a_D^{(0)} &=&-\frac{\pi}{\sqrt{2} } (u-1)  \, {}_2F_1\left(\frac{1}{2},\frac{1}{2};2;-\frac{1}{2} (u-1)\right)  \,.
\end{eqnarray}
Instead of evaluating the integrals over the coefficients $S_n$, $n \ge 0$ directly, one can define differential operators $D_{2n}$ of order $2n$ that map $S_{-1}$ to $S_{2n-1}$ up to total derivative terms \cite{Mironov:2009uv,He:2010if}. The first few of these are  \cite{He:2010if}
\begin{eqnarray}
D_{2} &=& \frac{1}{24}  (2 u \partial_u^2 + \partial_u) \,, \label{WKBpert} \\
D_{4} &=& \frac{1}{2^{7}} ( \frac{28}{45} u^2 \partial_u^4 + \frac{8}{3} u \partial_u^3 + \frac{5}{3} \partial_u^2) \,,
\nonumber \\
D_{6} &=& \frac{1}{2^{9}} (\frac{124}{945} u^3 \partial_u^6 + \frac{158}{105} u^2 \partial_u^5 + \frac{153}{35} u \partial_u^4 +
\frac{41}{14} \partial_u^3) \,,
\nonumber \\
D_{8} &=& \frac{1}{2^8} (\frac{127}{2^3 \times 4725} u^4 \partial_u^8 + \frac{13}{175} u^3 \partial_u^7 
+ \frac{517}{2^4 \times 63} u^2 \partial_u^6 + \frac{9539}{2^3 \times 945} u \partial_u^5 + \frac{15229}{2^7 \times 135} \partial_u^4) \nonumber \,.
\end{eqnarray}
Derivatives of the hypergeometric functions can in turn be rewritten as hypergeometric functions. By acting with the $D_{2n}$ on $a^{(0)}$ and $a_D^{(0)}$, one hence obtains the coefficients of the formal power series (\ref{WKB_periods}) again in terms of
hypergeometric functions of the modulus $u$. 

Note that expressing the higher order corrections to the $A$- and $B$-period via derivative operators acting on the leading contribution establishes that the monodromy matrix governing the behavior of $a^{(0)}$ and $a_D^{(0)}$ upon circling singularities in the $u$-plane is not corrected at any order in $\epsilon$: schematically, $D \left(a^{(0)} \log u \right) = D a^{(0)} \log u + a^{(0)} D \log u$, and $D \log u$ does not exhibit monodromy.

The two formal series allow us to 
eliminate $u$, and to solve for instance for ${\cal F} (a,\epsilon)$, which agrees with the perturbative expansion of the
Nekrasov instanton partition function.

We performed successful numerical checks on the perturbative WKB approximation in a range of parameters where the Stokes phenomena we shall discuss in the next subsection are numerically negligible. We give  an example of such a check
in appendix \ref{WKBpertnum}.

\subsection{Beyond perturbation theory} \label{bpt}
\label{beyondperturbationtheory}
We have seen that the monodromy of the solution to the null vector decoupling equation (\ref{Toda2}) plays an important role in our analysis: it corresponds to the exchanged momentum of the conformal block or, equivalently, to the scalar vacuum expectation value in the gauge theory. We also know by Floquet theory that a basis of exact solutions to the Mathieu equation (\ref{math_diff}), called Floquet solutions, exists with monodromy behavior $\psi_\pm \rightarrow e^{\pm \pi i \nu} \psi_\pm$. As we have reviewed above, the WKB ansatz gives rise to two independent formal solutions of the differential equation as a power series in $\epsilon$ which formally diagonalize the monodromy matrix; they hence approximate in a sense we shall discuss presently the Floquet solutions of the Mathieu equation. Their monodromies under $q \rightarrow q+ 2\pi i$ to leading order in $\epsilon$ are given by $\pm$ the $A$-period integral of $S_{-1}$, which can be identified with the Seiberg-Witten differential $\lambda$. This period hence provides an approximation to the characteristic exponent $\nu$ via $a \sim i \pi \epsilon \nu$. In this section, we will see how to incorporate $\exp[-1/\epsilon]$ corrections in this analysis. We introduce the notation $\aex$ for the $A$-period incorporating these corrections, such that $\aex=i \pi \epsilon \nu$. In terms of the quantities introduced in section \ref{irreg_block}, $\aex/\epsilon = \acft /\epsilon_1$.

The formal power series $\psi_{\pm}$ obtained from the WKB ansatz in the form (\ref{formal_WKB}) generically do not converge. Instead, they provide asymptotic expansions to actual solutions to the given differential equation: for $\epsilon$ contained in a sector $\arg \epsilon \in (\theta_1,\theta_2)$ (we will have much more to say about this range in the following), solutions $\Psi_{\pm}$ exist of the form $\Psi_{\pm} = \exp \left(\pm \frac{1}{\epsilon} \int^q \sqrt{Q_0(z)} \,dz \right) \epsilon^{1/2} f_{\pm}(q,\epsilon)$, with $f_\pm$ analytic in $q$ and in the given sector for $\epsilon$,
 such that for any $N\in \IN$ and $\rho>0$, a constant $C>0$ exists with
\be
|\epsilon|^{-N} |f_{\pm}(q,\epsilon) - \sum_{k=0}^{N-1} \psi_{\pm,k} \epsilon^k| \le C  \quad \forall \epsilon: |\epsilon|< \rho \,, \arg \epsilon \in (\theta_1,\theta_2)  \,.
\ee
It is easy to see from the definition that a function with an asymptotic expansion in terms of formal power series has precisely one such expansion. 

Borel resummation is a technique, given a formal power series in $\epsilon$, to construct a function analytic in a sector of the $\epsilon$-plane which has the formal series as its asymptotic expansion. The Borel sum is constructed in two steps. The Borel transform $\psi_B$ of a formal series $\psi$ is defined as
\be
\psi(\epsilon) = \sum_{k=0}^\infty \psi_k \epsilon^k \quad \rightarrow \quad \psi_B(\dualeps) = \sum_{k=1}^\infty \psi_k \frac{\dualeps^{n-1}}{(n-1)!}  \,.
\ee
If $\psi_B$ converges around $\dualeps = 0$ and can be analytically continued along a half-line $\ell_\theta$ connecting 0 to infinity at an angle $-\theta$ to the positive real $\dualeps$-axis, we can define the Laplace integral of $\psi_B$ in direction $\theta$ as
\be
\cS_\theta[\psi](\epsilon) = \psi_0 + \int_{\ell_\theta} e^{-\frac{\dualeps}{\epsilon}} \psi_B(\dualeps) \,d\dualeps \,.
\ee
If this integral exists, $\cS_\theta[\psi]$ provides the sought after analytic function; it is called the Borel sum of $\psi$ in direction $\theta$, and $\psi$ is called Borel summable. Notice that at given $\epsilon$, the existence of the integral generically constrains $\theta$ to lie within the sector 
\be \label{eps_theta}
\theta \in (- \frac{\pi}{2}-\arg \epsilon, \frac{\pi}{2} - \arg \epsilon) \,.
\ee
The WKB analysis of the Mathieu equation gives rise to formal WKB series that are assumed\footnote{The mathematical literature on this subject is uncharacteristically beset by assumptions and deferred proofs. But see \cite{MR3050812} for a proof of Borel summability in the case of a particular polynomial potential, and the forthcoming work \cite{Schaefke_to_appear} for the general polynomial case.} to be Borel summable away from a discrete infinite set of angles $\theta$. These angles partition the $\dualeps$-plane into sectors, half of which, given an $\epsilon$ and in accord with (\ref{eps_theta}), determine Borel sums of $\psi$, possibly differing by exponentially suppressed terms amongst each other. This ambiguity or integration path dependence of Borel resummation gives rise to the so-called Stokes phenomenon.

When the asymptotic series being resummed is the WKB solution to a differential equation, its coefficients depend on a parameter $q$, and Borel resummation under favorable circumstances leads to analytic solutions of the differential equation. For an ordinary differential equation of second order, this space is two-dimensional. The Stokes phenomena thus corresponds to assigning a different linear combination of a given basis of analytic solutions of the differential equation to WKB solutions via the process of Borel resummation.

When considering $q$-dependent coefficients, two types of singularities, mobile and fixed, can appear in the Borel plane; mobile singularities are those whose position depends on $q$. Away from isolated points in parameter space, all singularities that appear in the Borel plane are mobile. Our analysis will hence essentially focus on this case. Keeping the integration path of the Laplace transform in the Borel plane fixed, the Stokes phenomenon in this context manifests itself by discontinuities in the Borel resummation when $q$ crosses certain lines, called Stokes lines, on $\Sigma$. The Stokes lines divide $\Sigma$ into domains called Stokes regions. By the foregoing, the two solutions of the Mathieu equation that we obtain by Borel resummation of the formal WKB solutions depend on the Stokes region: the analytic continuation of the Borel resummation into a different Stokes region will equal a linear combination of the Borel resummed solutions native to that region. In other words, Borel resummation and analytic continuation in $q$ do not commute (before Borel resummation, analytic continuation is to be understood term by term).

Returning now to the question of determining the characteristic exponents of the Mathieu equation, we see that two phenomena need to be taken into account when passing from WKB to exact results. Firstly, the period of $S_{odd}$, which naively coincides with the characteristic exponent, must be Borel resummed. The resulting analytic function of $\epsilon$, called a Voros multiplier,\footnote{In the literature, this term is sometimes also used to indicate the formal period before Borel resummation.} will depend on the integration path chosen for the Laplace transform. Secondly, to determine the monodromy matrix of a pair of solutions requires analytically continuing them between different Stokes regions. As a consequence, we will see that the Floquet solutions to the differential equation do not coincide with the Borel resummation of the WKB solutions. Both manifestations of the Stokes phenomena must be taken into account to determine the characteristic exponents of the Mathieu equation. We will show that apparent ambiguities due to the choice of integration path cancel out in the process.

\subsubsection{The Stokes graphs}
\label{Stokespatterns}
Studying the Borel resummation behavior of the WKB solutions requires determining the Stokes graphs of the Mathieu equation (\ref{math_diff}) for a given choice of the parameters $u$ and $\epsilon$. These are entirely determined by the leading contribution $Q_0$ to the potential specifying the differential equation (\ref{gen_diff}). More precisely, upon a change of variables $q \rightarrow \tilde{q}$ and a rescaling of the solution $\Psi$ to absorb the ensuing first order derivative, (\ref{gen_diff}) remains form invariant upon replacing $Q_0(z)$ by
\be
\tilde{Q}_0(\tilde{z},\epsilon) = Q_0 (z(\tilde{z}) \left(\frac{d\tilde{z}}{d \tilde{z}} \right)^2 \,,
\ee
and shifting the higher order coefficients $Q_n$. The invariant quantity is hence
\be \label{quad_diff}
\phi = Q_0\, dz^{\otimes2} \,.
\ee
It is thus natural to interpret $Q_0$ as the coefficient of a section of the line bundle $K^{\otimes 2}$, with $K$ the canonical line bundle on $\Sigma$. Such sections are called quadratic differentials. We encountered them in the context of Seiberg-Witten theory in section \ref{sw_parametrization}. As we remarked above, the square root of the quadratic differential, $\lambda = S_{-1} \,dq$, coincides with the Seiberg-Witten 1-form of the underlying gauge theory; consequently, the quadratic differentials (\ref{quad_diff}) and (\ref{phi_sw}) coincide (up to irrelevant normalization). 

To render $\lambda$ single-valued, we introduce the double cover $\hat{\Sigma}$ of the Riemann surface $\Sigma$, branched at the simple zeros and poles (if present) of $\phi$. Note that since we have not restricted $\re q \ge 0$ in defining $\Sigma$, $\hat{\Sigma}$ does not coincide with the conventional Seiberg-Witten curve. In particular, the branch cuts in figure \ref{q_plane} in the upper and lower half-plane are not identified on $\hat{\Sigma}$, and a path connecting them does not yield a cycle. Since the two branch cuts can be chosen to be mapped into each other under $q \rightarrow -q$, and the formal power series $S_{odd}$ which determines the formal WKB series (\ref{sqrt_s_odd}) is even in $q$, we can essentially ignore this subtlety until the end of subsection \ref{the_core}. In an abuse of terminology, we will continue to refer to a path connecting the two turning points as the $B$-cycle or as homologous to the $B$-cycle. 
 
We define trajectories of the quadratic differential $\phi$, called WKB curves by \cite{Gaiotto:2009hg}, by the condition that $\lambda$ have constant phase along them. In other words, tangent vectors $\partial_t$ to trajectories satisfy
\begin{eqnarray}
e^{i \theta} \, \lambda \cdot \partial_t \in \mathbb{R}_+ \,.
\end{eqnarray}
This translates into the integral condition
\be \label{int_cond}
\im e^{i \theta} \int^q \lambda = \mbox{const} \,.
\ee
Distinguished points on the Riemann surface $\Sigma$ are given by the zeros of $\phi$. These are called turning points of the differential equation. Depending on the order of the zero, we distinguish between simple, double, or higher order turning points. A Stokes line is a trajectory that ends on a turning point $q_0$, hence satisfies the equation 
\be \label{def_stokes_line}
\im e^{i \theta} \int_{q_0}^q \lambda = 0 \,.
\ee
The graph formed by these Stokes lines is called the Stokes graph in the direction $\theta$.
As noted above, the relevance of Stokes lines stems from their relation to the position of the poles of the Borel transform of the formal WKB solutions: when $q$ lies on a Stokes line associated to an angle $\theta$, the Borel transform exhibits a pole on the line $\ell_{\theta}$.

For the analysis of transition behavior between Stokes regions, it is important to endow Stokes lines with an orientation. We will define the real part of $ e^{i \theta} \int^q \lambda$ to increase in the positive direction along a trajectory. This convention implies that a Stokes lines is oriented away from the turning point $q_0$ if $\re e^{i \theta} \int_{q_0}^q \lambda >0$ along it.

By equation (\ref{def_stokes_line}), we see that the pattern of Stokes lines is determined by $\theta \in [0,\pi)$; under $\theta \rightarrow \theta + \pi$, the pattern remains invariant, but the orientation of each Stokes line flips. 

A local analysis establishes that the number of Stokes lines emanating from a turning point is determined by its order, as follows: an order $n$ turning point leads to local behavior $z^{\frac{n}{2}+1}$ of the integral (\ref{int_cond}) and therefore has $n+2$ Stokes lines emanating from it, with angle $2\pi/(n+2)$ between two neighboring lines. Two types of Stokes lines will be relevant for our analysis: simple (or single or separating) Stokes lines, which run between a turning point and a pole of $\phi$, and double Stokes lines (or Stokes saddles) that run between two turning points. Simple Stokes lines that are oriented away from turning points will be called dominant, those oriented towards turning points recessive. The Stokes lines emanating from a turning point are alternately dominant and recessive, except upon crossing a branch cut. 

The Stokes graphs of the Mathieu differential equation have been studied in the Gaiotto-Witten variables in \cite{Gaiotto:2009hg}. We will study them on the $q$-cylinder more directly related to the traditional form of the Mathieu equation.
For the computation of non-perturbative corrections to the approximation to the characteristic exponent given by the $A$-period of the Seiberg-Witten differential, we will need to study the Stokes graphs for any complex value of the parameter pair $(u,\epsilon)$.

In our variables, the quadratic differential $\phi$ is given by
\be
\phi= (\cos q -u)\,dq \otimes dq  \,.
\ee
Its turning points on the cylinder lie at 
\be
q_{up/down} \in \cos^{-1} u \,,
\ee
with the subscript indicating the $q$-halfplane on which the respective preimage of $u$ lies. The differential $\phi$ has an essential singularity at the infinity of the complex plane. To analyze its behavior restricted to $\Sigma$, it is convenient to revert to the variable $z = \cos q$, to discover a pair of cubic poles, one at each point at infinity on $\Sigma$. The behavior at such poles is that trajectories which approach sufficiently are attracted to the pole along a tangent line (more generally, the number of tangent lines equals 2 less than the order of the pole) \cite{Bridge}.

\paragraph{Critical Stokes graphs}

Fixing a pair $(u,\epsilon)$, studying the occurrence of double Stokes lines at special angles $\theta$, called critical, is important for several reasons. We can imagine the $\dualeps$-plane\footnote{Recall that $\dualeps$ is the Borel dual variable to $\epsilon$.} being divided into sectors via the critical angles. The global topology of Stokes graphs is constant within each sector, and transitions upon crossing into a neighboring sector in a simple manner. Pairs of sectors related by reflection through the origin differ only in the orientation of all Stokes lines. Voros multipliers exhibit jumping behavior upon transition between sectors.

We can make qualitative statements about the occurrence of double Stokes lines by recourse to an observation of \cite{Klemm:1996bj,Gaiotto:2009hg} matching BPS states in the spectrum of the $\cN=2$ gauge theory associated to a given Seiberg-Witten curve with the occurrence of such lines. The charges of the BPS state correspond to the homology class associated to the line (recall that the double Stokes lines connect turning points on different sheets of the Riemann surface, hence determine closed curves on it). {From} our knowledge of the BPS spectrum of pure $\cN=2$ gauge theory \cite{Seiberg:1994rs,Ferrari:1996sv}, we are thus led to distinguish two regions on the $u$-plane, separated by the curve of marginal stability, which runs through the points $u= \pm 1$. Inside the curve, the BPS spectrum consists of the monopole and the dyon, of charge $\pm(0,1)$ and $\pm(1,\pm 1)$ respectively (the relative sign between the electric and magnetic charge of the dyon is not monodromy invariant). Hence, at fixed $\epsilon$ and any value of $u$ within this region, two values of the angle $\theta$ in the interval $ [0,\pi)$ should give rise to Stokes graphs exhibiting a double Stokes line. One of these will connect the turning points $q_{up}$ and $q_{down}$ directly, the other will wrap around the cylinder once before connecting the turning points. These angles shifted by $\pi$ correspond to the respective antiparticles. Outside the curve of marginal stability, the spectrum consists of infinitely many BPS particles, of charge $\pm (n,1)$ for $n \in \IZ$ (in addition to the vector bosons at charge $\pm(1,0)$, which also give rise to distinct Stokes patterns, see \cite{Klemm:1996bj,Gaiotto:2009hg}). At a given value of $u$ in this region, infinitely many values of $\theta$ in the interval $[0,\pi)$ hence give rise to double Stokes lines, one for each wrapping number $n \in \IZ$ around the cylinder.

We now turn to a more systematic study of the critical graphs.

\begin{figure}
\includegraphics[scale=1]{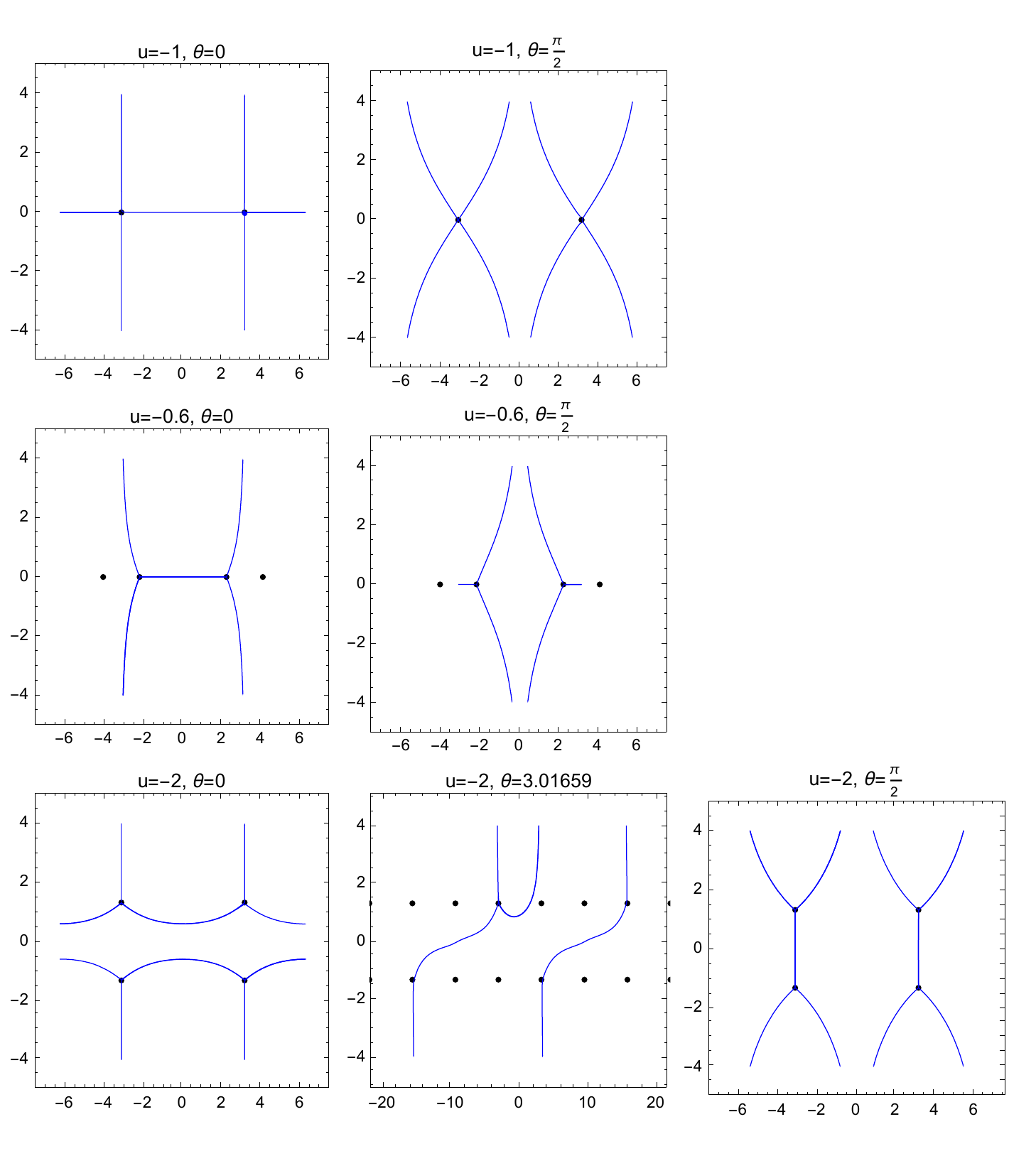}
\centering
\caption{Stokes graphs: the graph in the first row is for $u$ on the curve of marginal stability at $u=-1$, the second row depicts the two critical graphs inside the curve of marginal stability, and the third row a selection of critical graphs outside the curve of marginal stability.} \label{critical_graphs}
\end{figure}

\paragraph{Critical Stokes graphs at $u=\pm 1$} A good starting point for the systematic study of the Stokes graphs of the Mathieu equation is at the values of $u$ at which the quadratic differential exhibits double turning points, i.e. at the monopole and dyon points $u = \pm 1$. According to our general analysis above, $n+2=4$ Stokes lines emanate from each such point. For concreteness, let us consider the point $u=-1$. A critical Stokes graph for this choice of $u$ occurs at $\theta=0$, as $\lambda \cdot \partial_t \in \mathbb{R}$ along the line connecting the turning points at $q= -\pi$ and $q=\pi$ along the real axis. This double Stokes line corresponds to the monopole. The dyon at $u=-1$ is massless, the corresponding double Stokes line has zero length. It arises at $\theta = \frac{\pi}{2}$. The corresponding two graphs are depicted in the first row of figure \ref{critical_graphs}.

\paragraph{Critical Stokes graphs: inside the curve of marginal stability} Moving away from the singular points $u=\pm 1$ into the strong coupling region inside the curve of marginal stability, each double turning point splits into two single turning points. These have $n+2=3$ Stokes lines emanating from them. Keeping $u$ real, the two critical Stokes graphs still lie at $\theta=0$ and $\theta=\frac{\pi}{2}$, as depicted in the second row of figure \ref{critical_graphs}. The dyon acquires a mass, as the double Stokes line corresponding to $\theta= \frac{\pi}{2}$ now has finite length; along it, $\lambda$ is purely imaginary. Giving $u$ an imaginary part moves the turning points off of the real axis, while maintaining the topology of the diagrams. In particular, the two simple Stokes lines at each turning point run off to imaginary infinity in opposite half-planes.

\paragraph{Critical Stokes graphs: outside the curve of marginal stability}

Starting from $u=-1$, moving into the weak coupling region outside the curve of marginal stability while keeping $u$ real gives rise to the Stokes graphs depicted in the third row of figure \ref{critical_graphs}. The double Stokes lines in the leftmost diagram correspond to the $n\rightarrow \infty$ limit of the BPS particles of charge $n\, a + a_D$. Within a small interval around $\theta=0$, an infinite number of double Stokes lines arise, which wrap, as $\theta$ approches 0, an increasing number $n$ of times around the cylinder before connecting the two turning points, corresponding to central charge $n\, a+ a_D$. The second graph in the last row of figure \ref{critical_graphs} corresponds to the value $n=2$. Moving away from $\theta=0$, $n$ decreases, till it reaches 0 at $\theta = \frac{\pi}{2}$, as depicted in the final graph in the third row. Further increasing $\theta$ yields double Stokes lines that wrap the cylinder in the opposite direction. Giving $u$ an imaginary part shifts the turning points away from the line $\re q =\pm \pi$, while maintaining the topology of the diagram. In particular, aside from the two critical graphs corresponding to $n \rightarrow \infty$ with two double Stokes lines attached to each turning point, all other critical graphs exhibit two simple Stokes lines at each turning point moving off to imaginary infinity in the same half-plane.

\vspace{0.5cm}

In accord with the BPS analysis above, the $\dualeps$-plane is hence split into four sectors inside the curve of marginal stability. Outside this curve, it is split into infinitely many sectors that accumulate at $\theta=0$ and $\theta= \pi$. We will introduce a convenient indexation of these sectors below.

\begin{figure}
\includegraphics[scale=1]{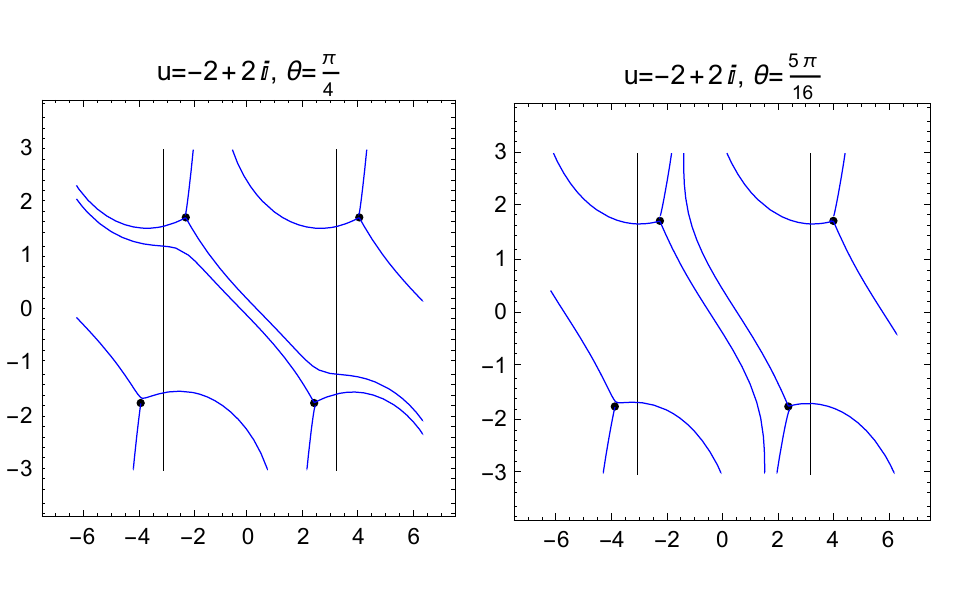}
\centering
\caption{Two Stokes graphs related by a flip. If $\theta_0$ denotes the critical angle, and $\delta$ is a small positive constant, the graph to the left is representative of the topology at $\theta_0 - \delta$, the graph to the right of that at $\theta_0 + \delta$.} \label{TheFlip}
\end{figure}

\paragraph{Generic Stokes graphs} As we move off of a critical value of $\theta$, the double Stokes line $\ell_0$ splits into two simple Stokes lines. These can avoid each other in two topologically distinct manners: upon decreasing $\theta$ away from a critical value, the lines swerve to the left of $\ell_0$ as seen from the turning point from which they emerge, upon increasing $\theta$, they swerve to the right. This behavior is visible in figure \ref{TheFlip} for $u$ lying outside the curve of marginal stability, and again in figure \ref{ICMO} for $u$ inside the curve. The Stokes graphs on the two sides of the critical $\theta$-value are said to be related via a flip.

The generic Stokes graph hence has three Stokes lines emerging from each turning point and running off towards infinity, two towards infinity in the upper half-plane of the $q$-plane and one towards infinity in the lower half-plane, or vice versa. Inside the curve of marginal stability, these two cases are interchanged by a flip. Outside this curve, the turning point in the upper half-plane always exhibits two simple Stokes lines running off to infinity in the positive half-plane. By the symmetry $q \rightarrow -q$ of the equation, the behavior at the alternative turning point is obtained by reflection through the origin. 

\subsubsection{Computing non-perturbative corrections to the characteristic exponent} \label{the_core}
Having determined the Stokes graphs of the Mathieu equation, we are now in a position to compute contributions to the monodromy of its solutions that are not visible via formal WKB analysis.

As explained in the introduction to this section, the Borel resummations of the two formal WKB solutions $\psi_{\pm}(q,\epsilon)$, at fixed $\epsilon$, yield a different basis of solutions of the Mathieu equation depending on the Stokes region in which the argument $q$ lies. The solutions $\Psi_\pm^{(i)}$ obtained upon Borel resummation with $q$ in Stokes region $(i)$ can be analytically continued into a neighboring Stokes region $(j)$, yielding a basis of solutions also here. A second such basis can be obtained directly by Borel resummation of $\psi_{\pm}(q,\epsilon)$ with $q$ chosen in region $(j)$. The matrix $S_{(i) \rightarrow (j)}$ relating these two bases of solutions is referred to as the connection matrix from region $(i)$ to region $(j)$. It depends on the choice of normalization of the WKB solutions. In this paper, we will, as pointed out in section \ref{MathieuWKB}, normalize the WKB solutions $\psi_{\pm}$ at the turning point from which the Stokes line emanates that we wish to cross. Denoting the turning points as $q_k$, we will use the notation $(q_k)_{\pm}^{(i)}$ to indicate the Borel resummations in Stokes region $(i)$ of the WKB solutions normalized at turning point $q_k$, with $q_k$ lying on a boundary of the Stokes region $(i)$.

In this notation, 
\be
\left( \begin{array}{c} (q)_+^{(i)}  \\ (q)_-^{(i)}  \end{array} \right)(q^{(j)}) =  S_{(i) \rightarrow (j)} \left( \begin{array}{c} (q)_+^{(j)}  \\ (q)_-^{(j)} \end{array} \right)  (q^{(j)}) \,, \label{an_cont_dom}
\ee
where $q^{(j)}$ labels a point in Stokes region $(j)$ neighboring Stokes region $(i)$.  With the given choice of normalization, the matrices $S_{(i) \rightarrow(j)}$ take a simple form (\cite{Voros}, theorem 2.25 of \cite{IN}): upon analytically continuing across a dominant Stokes line counterclockwise with regard to the turning point, it is given by \cite{IN}
\be  \label{st_dominant}
S_{dom} =\left(\begin{array}{cc} 1 & i \\ 0 & 1 \end{array} \right)  \,,
\ee
whereas analytic continuation counterclockwise across a recessive Stokes line, requires the connection matrix \cite{IN}
\be   \label{st_recessive}
S_{rec} = \left(\begin{array}{cc} 1 & 0 \\ i & 1 \end{array} \right) \,.
\ee
As a consistency check, note that a full revolution around a turning point yields the unit matrix as connection matrix:
\be \label{trans_cons}
S_{rec} S_{branch} S_{rec} S_{dom} = \left(\begin{array}{cc} 1 & 0 \\ 0 & 1 \end{array} \right) \,, \quad  S_{dom} S_{branch} S_{dom} S_{rec} = \left(\begin{array}{cc} 1 & 0 \\ 0 & 1 \end{array} \right) \,,
\ee
with
\be
S_{branch} = -i \left(\begin{array}{cc}  0 & 1 \\ 1 & 0 \end{array} \right) 
\ee
the transition matrix upon crossing a branch cut. The factor $-i$ is due to the square root in the denominator of (\ref{sqrt_s_odd}).

{From} the form of the WKB solution (\ref{WKB_general_form}), we can read off the following relation between two solutions $(q_1)_\pm^{(i)}$ and $(q_2)_\pm^{(i)}$, when $q_1$ and $q_2$ lie on the boundary of the same Stokes region $(i)$:
\be
(q_1)_{\pm}^{(i)} =   \left( \exp\left[\pm \int_{q_1}^{q_2} S_{odd} \right] \right)_s (q_2)_{\pm}^{(i)}  \,.
\label{shift}
\ee
The prefactor of $(q_2)_{\pm}^{(i)} $ is the Voros multiplier associated to the cycle represented by the line connecting the two turning points $q_1$ and $q_2$: the notation $(\cdot)_s$ denotes the Borel resummation of the formal power series in parentheses. The subscript $s$ indicates the sector in the $\dualeps$-plane in which this resummation is performed: as we will review below, Voros multipliers are locally constant functions of $\theta$ in a given sector; they can jump as $\theta$ crosses a critical angle. 
We have suppressed the sector dependence elsewhere in the notation.

\begin{figure}
\includegraphics[scale=0.6]{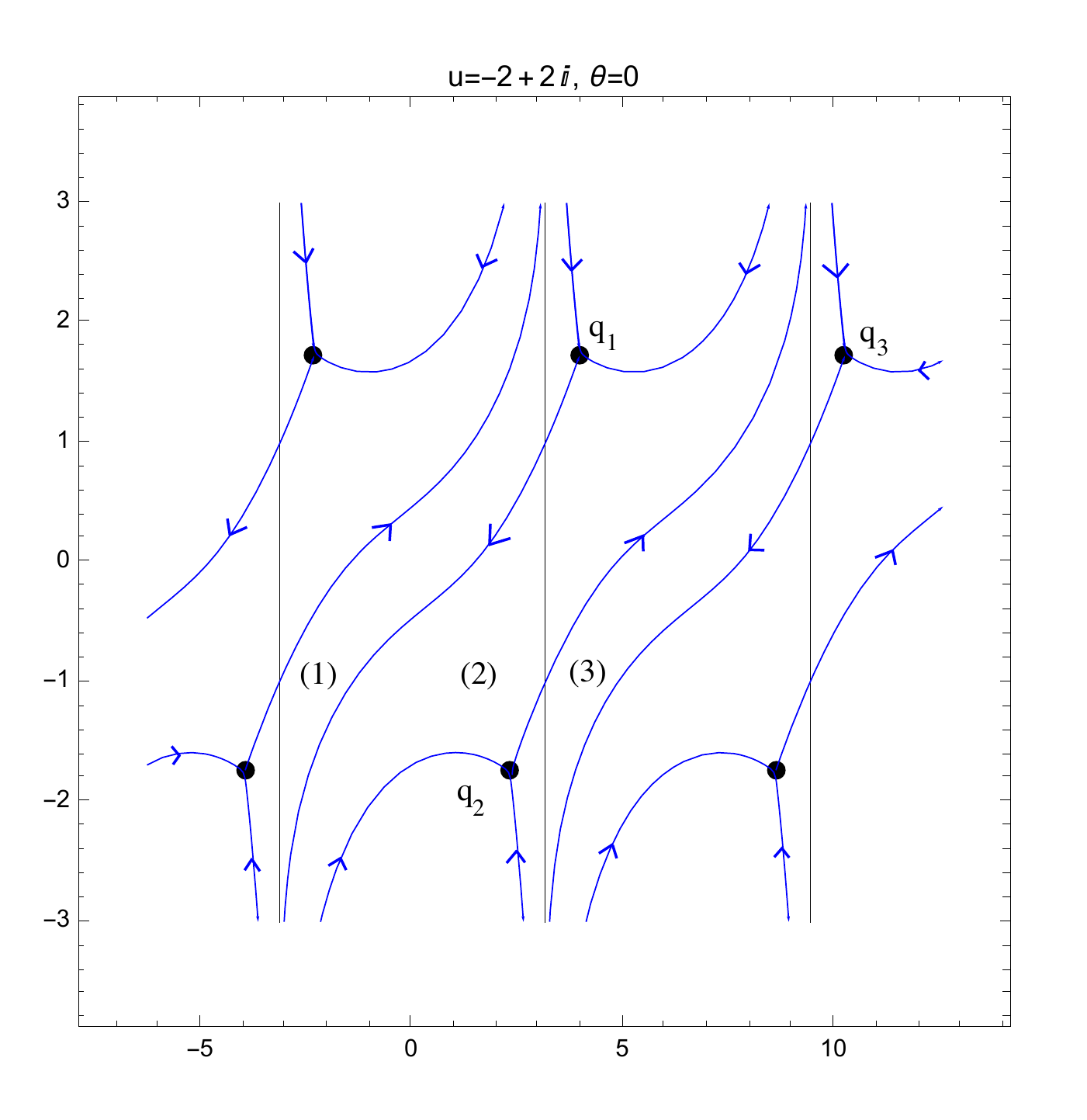}
\centering
\caption{The Stokes regions (1), (2), (3) and turning points $q_1$, $q_2$, $q_3$ as introduced in the text.} \label{transition_graph}
\end{figure}

\paragraph{The $A$-monodromy outside the curve of marginal stability}
Computing the $A$-monodromy for any value of the parameter pair $(u,\epsilon)$ with $u$ outside the curve of marginal stability, requires crossing at least two Stokes lines. To standardize our calculations, we will always start off at a point $q$ in a Stokes region, henceforth Stokes region (1), chosen such that the first Stokes line to cross in analytically continuing the solution along the negative $A$-cycle (recall from section \ref{sw_parametrization} that the $A$-cycle runs from $\pi$ to $-\pi$ in the $q$-plane) is a Stokes line connected to a turning point in the upper half-plane, $q_1$. The analytic continuation across this line into Stokes region (2) will involve one of the two connection matrices (\ref{st_dominant}) or (\ref{st_recessive}), depending on whether the Stokes line is dominant or recessive. The next Stokes line to cross is a Stokes line emanating from a turning point, $q_2$, in the lower half-plane, and has opposite orientation. We thus arrive in Stokes region (3), which is identified with Stokes region (1) by the periodicity of the problem. Choosing our branch cuts, as we shall do throughout, to connect the turning points in the upper/lower half plane to imaginary infinity in the same half plane (see figure \ref{q_plane}), no branch cut is crossed along this path of analytic continuation. The terminology introduced here is exemplified in figure \ref{transition_graph}.

The terminology introduced in the previous paragraph allows us to introduce the indexing of sectors of the $\dualeps$-plane promised above: given $q_i \in [-\pi + 2\pi k_i, \pi + 2 \pi k_i]$ for $k_{1,2} \in \IZ$, we define $n=k_1 - k_2$, such that the cycle connecting $q_1$ and $q_2$ is homologous to $\pm(nA +B)$, with the orientation of the cycle chosen to coincide with that of the double Stokes line connecting the two turning points at the appropriate boundary of the $\theta$-sector. The sign is correlated with whether the first Stokes line crossed is dominant $(+)$ or recessive $(-)$. At a given choice of $(u,\epsilon)$, the sectors of the $\dualeps$-plane are hence uniquely indexed by $(n,\pm)$. If an angle $\theta$ lies within the sector $(n,\pm)$, then $\theta + \pi$ yields the same Stokes graph with the orientation of all Stokes lines reversed, hence lies in the sector $(n,\mp)$. In accord with (\ref{eps_theta}), exactly one of these choices of $\theta$ is compatible with the phase of the parameter $\epsilon$, which we have kept fixed throughout the argument. Flipping the sign of $\epsilon$ necessitates making the alternative choice. In this sense, flipping the sign of $\epsilon$ results in the map $(n,\pm) \mapsto (n,\mp)$,
\be \label{sign_flip_eps}
\epsilon \mapsto - \epsilon \quad   \Rightarrow \quad  (n,\pm) \mapsto (n,\mp) \,.
\ee

Introducing the notation $N_{i \rightarrow j}$ for the matrix mapping a basis of solutions obtained via Borel resummation of formal WKB functions normalized at turning point $q_i$ to the basis obtained from WKB functions normalized at turning point $q_j$, the analytic continuation we have described is encapsulated in the following equation:
\begin{eqnarray}  \label{monodromy_eq}
\left( \begin{array}{c} (q_1+2\pi)_+^{(3)} \\ (q_1+2\pi)_-^{(3)} \end{array} \right) (q+ 2\pi) = N_{2 \rightarrow 3}  S_{(2) \rightarrow (3)}^{-1}  N_{1 \rightarrow 2}  S_{(1) \rightarrow (2)}^{-1}\left( \begin{array}{c} (q_1)_+^{(1)} \\ (q_1)_-^{(1)} \end{array} \right) (q+2\pi) \,.
\end{eqnarray}
By the periodicity of the differential equation,
\be 
\left( \begin{array}{c} (q_1+2\pi)_+^{(3)} \\ (q_1+2\pi)_-^{(3)} \end{array} \right) (q+ 2\pi) = \left( \begin{array}{c} (q_1)_+^{(1)} \\ (q_1)_-^{(1)} \end{array} \right) (q) \,,
\ee
thus permitting us to identify the product of matrices in (\ref{monodromy_eq}) as the monodromy matrix $M_A$ (recalling again that the $A$-cycle runs from $\pi$ to $-\pi$). With the notation introduced in subsection \ref{M_Floquet} and by Floquet theory,
\be
\tr M_A = e^{\frac{\aex}{\epsilon}} + e^{-\frac{\aex}{\epsilon}} \,.
\ee
With the conventions introduced above, the appropriate normalization matrices $N_{i \rightarrow j}$ are given by
\begin{eqnarray}
 N_{1 \rightarrow 2}  &=& \left( \begin{array}{cc}  e^{-\frac{a_D+n\,a}{\epsilon}} & 0 \\ 0 &  e^{\frac{a_D+ n\, a}{\epsilon}}  \end{array} \right)_s \,,
\end{eqnarray}
and
\begin{eqnarray}
 N_{2 \rightarrow 3}  &=& \left( \begin{array}{cc}  e^{\frac{a_D+(n+1)\,a}{\epsilon}} & 0 \\ 0 & e^{-\frac{a_D+(n+1)\,a}{\epsilon}}  \end{array} \right)_s  \, .
\end{eqnarray}
The formal periods $a$ and $a_D$ were introduced in (\ref{WKB_periods}), and the subscript $s$ denotes the sector in which the Borel resummation is to be performed.
 
Substituting all matrices into equation (\ref{monodromy_eq}), we obtain the following trace of the monodromy matrix:
\be
\tr M_A = \left( 2 \cosh \frac{a}{\epsilon} + e^{\mp\frac{1}{\epsilon}(a(1+2n) + 2 a_D)} \right)_{(n,\pm)} \label{finaltrace_out}
\ee
At a given choice of the parameter pair $(u,\epsilon)$, a unique answer for $\tr M_A$ exists -- in particular, this answer cannot depend on the choice of integration direction $\theta$ of the Laplace transform, i.e. the choice of sector $(n,\pm)$. We hence need to explain the two apparent $\theta$-dependencies of this result: that we have a priori arrived at two different expressions, depending on whether the first Stokes line crossed is dominant or recessive, i.e. whether $\theta$ lies in the sector $(n,+)$ or $(n,-)$, and the $n$-dependence of these expressions. The key to resolving the first apparent ambiguity lies in the relation (\ref{eps_theta}): the phase of $\epsilon$ determines whether the sector $(n,+)$ or $(n,-)$ is appropriate. Moreover, by (\ref{sign_flip_eps}), flipping the sign of $\epsilon$ leaves the result invariant. To address the apparent $n$ dependence of the result, we will need to discuss the jumping behavior for Voros multipliers. We will do this in subsection \ref{voros_jumps} below.

\begin{figure}
\includegraphics[scale=1]{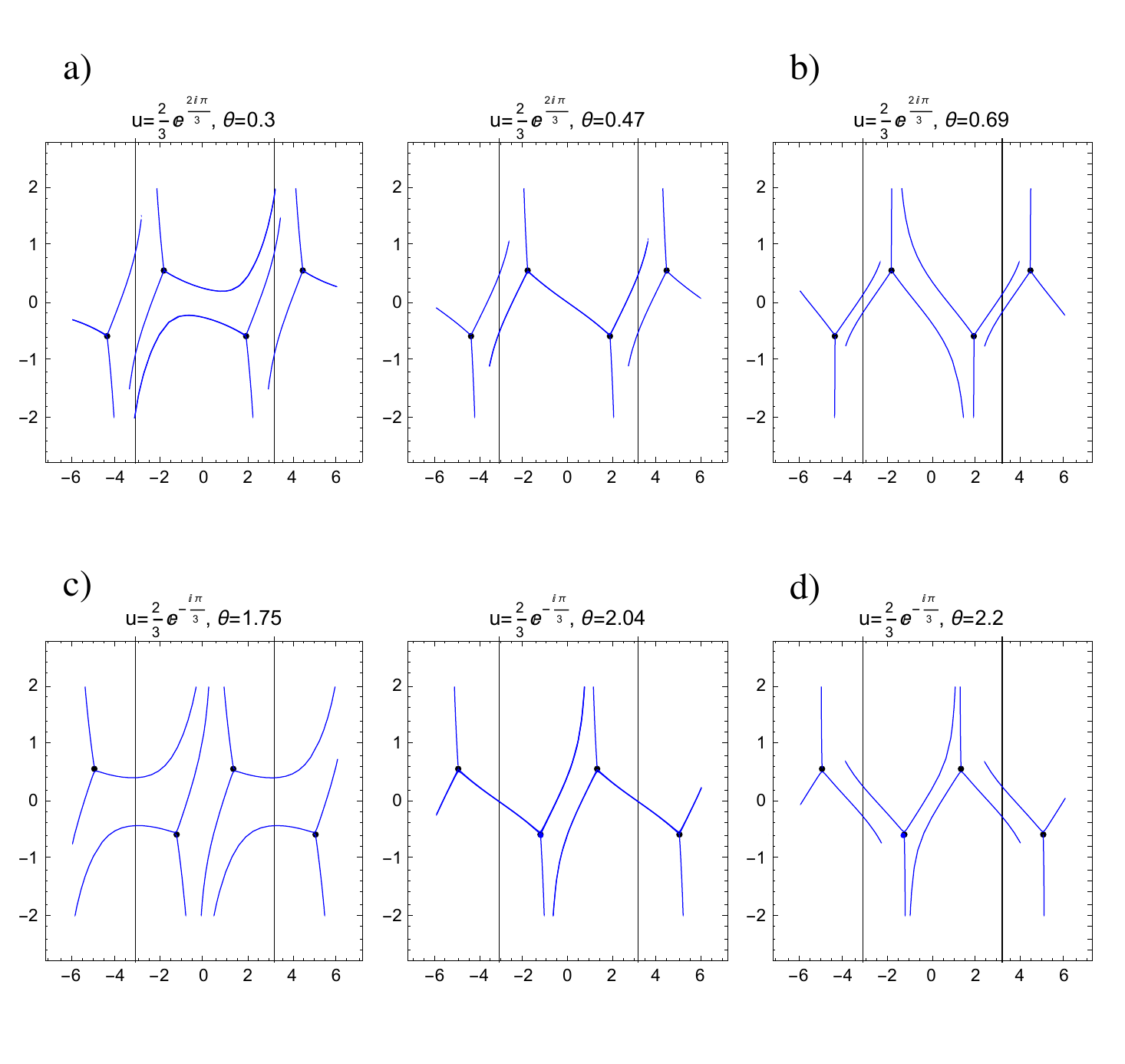}
\centering
\caption{Inside the curve of marginal stability.} \label{ICMO}
\end{figure}

\paragraph{The $A$-monodromy inside the curve of marginal stability} Representative graphs for the two sectors (up to orientation inversion) that arise inside the curve of marginal stability are depicted in figure \ref{ICMO}. 
The graphs in the first row arise at $\im u>0$, those in the second at $\im u<0$. The topology of the graphs \ref{ICMO}a/\ref{ICMO}c can be distinguished from that of the graphs \ref{ICMO}b/\ref{ICMO}d by whether the turning point in the upper half-plane is connected to imaginary infinity in the upper half-plane via one or two Stokes lines. We introduce the notation $(+--)$ and $(++-)$ to distinguish the corresponding $\theta$-sectors. The graphs that arise in the sector $(++-)$ for $\im u>0$, $\im u<0$ match the graphs in the sectors $(0,\pm)$, $(-1,\pm)$ outside the curve of marginal stability. The corresponding trace of the monodromy matrix thus follows from (\ref{finaltrace_out}) and is given by
\be
\tr M_A = \begin{cases}
\left( 2 \cosh \frac{a}{\epsilon} + e^{\pm \frac{1}{\epsilon}(a + 2 a_D)} \right)_{(++-)}
& \text{if }\im u >0 \quad \text{(graph \ref{ICMO}a),} \\
\left(2 \cosh \frac{a}{\epsilon} + e^{\pm\frac{1}{\epsilon}(-a + 2 a_D)} \right)_{(++-)}
& \text{if } \im u <0 \quad \text{(graph \ref{ICMO}c).}
\end{cases} \label{tr2ICMS}
\ee
To compute the $A$-monodromy for the sector $(+--)$, we can either cross four Stokes lines, or cross two Stokes lines and two branch cuts. The two computations are related by the relation (\ref{trans_cons}). The former choice translates, in the case that the first Stokes line crossed is dominant, into the sequence of transition matrices
\be
N_{2 \rightarrow 3} S_{rec} S_{dom} N_{1 \rightarrow 2} S^{-1}_{rec} S^{-1}_{dom} \,,
\ee
with
\be
N_{1 \rightarrow 2}  = \left( \begin{array}{cc}  e^{-\frac{a_D}{\epsilon}} & 0 \\ 0 &  e^{\frac{a_D}{\epsilon}}  \end{array} \right) \,, \quad N_{2 \rightarrow 3} = \left( \begin{array}{cc}  e^{\frac{a_D+a}{\epsilon}} & 0 \\ 0 &  e^{-\frac{a_D+  a}{\epsilon}}  \end{array} \right)
\ee
for \ref{ICMO}b and
\be
N_{1 \rightarrow 2}  = \left( \begin{array}{cc}  e^{\frac{a_D}{\epsilon}} & 0 \\ 0 &  e^{-\frac{a_D}{\epsilon}}  \end{array} \right) \,, \quad N_{2 \rightarrow 3} = \left( \begin{array}{cc}  e^{\frac{a_D-a}{\epsilon}} & 0 \\ 0 &  e^{-\frac{a_D-  a}{\epsilon}}  \end{array} \right)
\ee
for \ref{ICMO}d. The transition matrices upon shifting $\theta$ by $\pi$ are obtained by exchanging $S_{dom}$ and $S_{rec}$, which results in changing the sign of both $a$ and $a_D$ in the monodromy matrix. We thus obtain
\be
\tr M_A = \begin{cases}
\left( 2 \cosh \frac{2 a_D+a}{\epsilon} + e^{\pm \frac{1}{\epsilon}a} \right)_{(+--)}
& \text{if }\im u >0 \quad \text{(graph \ref{ICMO}b),} \\
\left(2 \cosh \frac{2a_D-a}{\epsilon} + e^{\pm\frac{1}{\epsilon}a} \right)_{(+--)}
&  \text{if } \im u <0 \quad \text{(graph \ref{ICMO}d).}
\end{cases} \label{tr4ICMS}
\ee
Upon inspection of figure \ref{ICMO}, one concludes that the results for $\im u>0$ should be mapped to those for $\im u<0$ via the map $a_D \mapsto a_D -a$; this relation is in accord with the results of our computation.

As was the case outside the curve of marginal stability, we again obtain multiple results for the trace of the monodromy matrix at a given fixed parameter $u$. The orientation dependence as reflected in the $\pm$ in the exponents in (\ref{tr2ICMS}) and (\ref{tr4ICMS}) is again resolved upon fixing the parameter pair $(u,\epsilon)$: only one orientation is compatible with a given choice of $\epsilon$ due to the relation (\ref{eps_theta}). We will see in section \ref{voros_jumps} that the apparent difference between the results for the sectors $(++-)$ and $(+--)$ is accounted for by the jumping behavior of Voros multipliers.

\paragraph{The $B$-monodromy} From the point of view of the Seiberg-Witten curve, a natural next task is to determine the $B$-monodromy of the solutions to the differential equation (\ref{math_diff}). However, from the point of view of the $q$-plane on which the differential equation is formulated, the notion of $B$-monodromy is not meaningful, as the turning points $q_{up/down}$ are not identified. The best we can do is ask how the wave functions obtained from Borel resummation of the WKB solution are related at $q_{up/down}$:
\ba
\psi_{\pm}(q_{down})\rightarrow \psi_{\pm}(q_{up})& =& \frac{1}{\sqrt{S_{odd}(q_{up})}} \exp \left(\pm \int^{q_{up}} S_{odd}(q) \, dq \right) \\ &=&\frac{1}{\sqrt{S_{odd}(q_{down})}} \exp \left(\pm (\int_{q_{down}}^{q_{up}} + \int^{q_{down}}) S_{odd}(q) \, dq \right)  \,.
\ea
In the final step, we have invoked the parity of the formal series $S_{odd}$. As both inside and outside the curve of marginal stability, an angle $\theta$ exists at which we can connect $q_{up}$ and $q_{down}$ without crossing a Stokes line, we can elevate this relation to the level of the Borel resummed functions:
\be \label{b_of_sorts}
\Psi_\pm (q_{down}) \rightarrow \left(e^{\pm\frac{a_D}{\epsilon}}\right)_s \Psi_\pm (q_{down})  \,,
\ee
where the arrow indicates transport along a path connecting $q_{down}$ to $q_{up}$. On the Seiberg-Witten curve $\Sigma/\!\!\sim$, the quotient given by the identification $q \rightarrow -q$ along the branch cut, equation (\ref{b_of_sorts}) indeed describes the $B$-monodromy of $\Psi_{\pm}$.

\subsubsection{Jumping phenomena for Voros multipliers}  \label{voros_jumps}
It is possible to choose the  normalization of WKB solutions such that their Borel transform remains locally invariant even as $\theta$ traverses a critical value \cite{IN}. This result implies that Voros multipliers must be sector dependent, and allows the determination of their transition behavior \cite{IN}. Let $\theta_0$ denote a critical angle at which a Stokes graph exhibits a single double Stokes line $\ell_0$. Given a cycle $\gamma$ on the Riemann surface $\hat{\Sigma}$ (the double cover of $\Sigma$, see discussion below \ref{quad_diff}), the associated Voros multipliers $(e^{\frac{a_\gamma}{\epsilon}})_-$ and $(e^{\frac{a_{\gamma}}{\epsilon}})_+$ at $\theta_0-\delta$ and $\theta_0+\delta$, $\delta$ a small positive constant, are related as follows \cite{IN}:
\be  \label{jump}
(e^{\frac{a_\gamma}{\epsilon}})_- = (e^{\frac{a_\gamma}{\epsilon}})_+ (1 + (e^{\frac{a_{\gamma_0}}{\epsilon}})_+)^{-(\gamma_0,\gamma)} \,.
\ee
Here, $\gamma_0$ is a cycle on $\hat{\Sigma}$ whose projection onto $\Sigma$ encircles $\ell_0$, with orientation chosen such that
\be \label{normalization_gamma}
\re e^{i \theta} \oint_{\gamma_0} \lambda \, dz < 0 \,.
\ee
The intersection pairing $(\cdot,\cdot)$ is chosen such that upon projection on $\Sigma$, the real and imaginary $q$-axis have intersection number one.

\paragraph{Outside the curve of marginal stability} Above, we labelled the sectors in between critical angles by an integer $n$ determining the topology of the Stokes graph, and a sign determining orientation. The integer $n$ increases with $\theta$, as we will argue in appendix \ref{local_analysis}. The double Stokes line $\ell_{(n,\pm)}$ which occurs at an angle $\theta$ on the boundary between sectors $(n-1,\pm)$ and $(n,\pm)$ is homologous to the cycle $\pm(B+nA)$, and therefore the cycle $\gamma_{(n,\pm)}$ encircling it, with the orientation choice given by (\ref{normalization_gamma}), is homologous to $\mp(2B + 2nA)$. By the property (\ref{jump}), the Voros multiplier associated to this cycle does not jump at the splitting of the Stokes line, i.e.
\be
(e^{\frac{2}{\epsilon}(a_D + n a)})_{(n,\pm)}=(e^{\frac{2}{\epsilon}(a_D + n a)})_{(n-1,\pm)} \,.
\ee
Noting finally that 
\be
-(\gamma_{(n,\pm)},\mp \gamma_A) = 1 \,,
\ee
with $\gamma_A$ of $\hat{\Sigma}$ a simple cover of the $A$-cycle on $\Sigma$, we can establish the independence of the trace of the monodromy matrix from the sector in which $\theta$ lies as follows:
\ba
\tr M_A &=& \left( e^{\frac{a}{\epsilon}}+e^{-\frac{a}{\epsilon}} + e^{\mp\frac{1}{\epsilon}(a(1+2n) + 2 a_D)}\right)_{(n,\pm)} \\
&=& \left( e^{\mp \frac{a}{\epsilon}} \right)_{(n,\pm)} \left( 1 + e^{\mp \frac{1}{\epsilon}(2n a + 2 a_D)}\right)_{(n,\pm)} + \left(e^{\pm \frac{a}{\epsilon}}\right)_{(n,\pm)} \\
&=& \left( e^{\mp \frac{a}{\epsilon}} \right)_{(n,\pm)} \left( 1 + e^{\frac{1}{\epsilon}a_{\gamma_{(n,\pm)}}}\right)_{(n,\pm)}^{-(\gamma_{(n,\pm)},\mp \gamma_A)} + \left(e^{\pm \frac{a}{\epsilon}}\right)_{(n,\pm)} \\
&=& \left( e^{\mp \frac{a}{\epsilon}} \right)_{(n-1,\pm)} + \left(e^{\pm\frac{a}{\epsilon}}\right)_{(n-1,\pm)} \left( 1 + e^{\frac{1}{\epsilon} a_{\gamma_{(n,\pm)}}  }\right)_{(n,\pm)}^{-(\gamma_{(n,\pm)},\mp \gamma_A)} \\
&=& \left( e^{\mp \frac{a}{\epsilon}} \right)_{(n-1,\pm)} + \left(e^{\pm\frac{a}{\epsilon}}\right)_{(n-1,\pm)} \left( 1 + e^{\frac{1}{\epsilon} a_{\gamma_{(n,\pm)}}  }\right)_{(n-1,\pm)}^{-(\gamma_{(n,\pm)},\mp \gamma_A)} \\
&=& \left( e^{\pm\frac{a}{\epsilon}}  + e^{\mp \frac{a}{\epsilon}} + e^{\mp \frac{1}{\epsilon}(a(1+2(n-1)) + 2 a_D)}\right)_{(n-1,\pm)} \,.
\ea

\paragraph{Inside the curve of marginal stability} Let us discuss the transition from the sector $(++-)$ to $(+--)$ at $\im u>0$ (figure \ref{ICMO}a to \ref{ICMO}b). The one at $\im u<0$ will then follow upon the mapping $a \mapsto a_D -a $. The cycle $\gamma_0$ in this case is given by $\gamma_0 = \mp 2B$, where the $-$ sign is for the case that the first line crossed in \ref{ICMO}a is dominant. It follows that
\be
\left( e^{\mp \frac{a}{\epsilon}} \right)_{(++-)} = \left( e^{\mp \frac{a}{\epsilon}} \right)_{(+--)} ( 1 + e^{\mp \frac{2a_D}{\epsilon}}) \,,
\ee
and hence
\ba
\tr M_A &=& \left( e^{\frac{a}{\epsilon}} + e^{-\frac{a}{\epsilon}} + e^{\pm \frac{a+ 2 a_D}{\epsilon}} \right)_{(++-)} = \,\, \left( e^{\pm \frac{a}{\epsilon}} (1 + e^{\pm \frac{2 a_D}{\epsilon}})  + e^{\mp\frac{a}{\epsilon}} \right)_{(++-)}  \\
&=& \frac{\left( e^{\pm \frac{a}{\epsilon}} \right)_{(+--)} (1 + e^{\pm \frac{2 a_D}{\epsilon}})}{  1 + e^{\mp\frac{2a_D}{\epsilon}}} +  \left( e^{\mp \frac{a}{\epsilon}} \right)_{(+--)} ( 1 + e^{\mp \frac{2a_D}{\epsilon}}) \\
&=& \left( e^{\frac{2a_D+a}{\epsilon}} + e^{-\frac{2a_D+a}{\epsilon}} + e^{\mp \frac{a}{\epsilon}} \right)_{(+--)} \,.
\ea

\subsubsection{At the singular points $u=\pm1$}
 The $\epsilon$-neighborhoods of the singular points $u=\pm 1$ have received particular attention in the literature \cite{ZinnJustin:2004ib,ZinnJustin:2004cg,Delabaere,Gorsky:2014lia,Basar:2015xna,Krefl:2014nfa}, as one is driven to these points at small real $\epsilon$ and constant real characteristic exponent $\nu$ (see the discussion around equation (\ref{asym_small_eps})). {From} the gauge theory perspective, $u=\pm 1$ are the points on moduli space where an extra state becomes massless and the
 effective gauge theory description breaks down. Nevertheless, we briefly touch upon this region in this subsection, and show how existing results for $\tr M_A$ align with those we found above.

The analysis presented in subsection \ref{the_core} must be modified for the choice  of modulus $u= \pm 1 + \epsilon\, U$, as the Stokes analysis depends on the function $Q_0$, the leading term of $Q(q,\epsilon)$ as introduced in (\ref{gen_diff}), and is hence blind to the $\epsilon\, U$ distance from the singular points. At these points, the two turning points of $Q_0$ coalesce into a double turning point, and the connection matrices (\ref{st_dominant}) and (\ref{st_recessive}) are no longer valid. A calculation in the same spirit as the ones presented in subsection \ref{the_core} was performed in \cite{Delabaere} at these points, proving a conjecture of \cite{ZinnJustin:2004ib,ZinnJustin:2004cg} for the form of the characteristic exponents for this choice of $u$. The expression for $\tr M_A$ as cited e.g. in equation (2.14) of \cite{Basar:2015xna} for $u=-1$ can be expressed in terms of the periods $a$ and $a_D$ by invoking the formulas (3.28), (3.8), and (3.33) of \cite{Basar:2015xna}, yielding
\ba
\tr M_A
&=&  e^{\frac{1}{\epsilon} (a + 2 a_D) } + e^{-\frac{1}{\epsilon}a} + e^{\pm \frac{1}{\epsilon}(a+a_D) - \frac{a_D}{\epsilon}} \\
&=& \begin{cases}
2\cosh \frac{a+2a_D}{\epsilon}+e^{-\frac{1}{\epsilon}a} \\
2\cosh \frac{a}{\epsilon} + e^{\frac{1}{\epsilon}(a+2a_D)} \, ,
\end{cases}  
\label{exactWKBneardouble}
\ea
with the periods $a$ and $a_D$ evaluated at $u=-1 + \epsilon \, U \in [-1,1]$. This result coincides with the characteristic exponents for $\im u>0$ inside the wall of marginal stability found above, for a particular orientation of the Stokes graphs. The two signs $\pm$ in the first line of this equation are due to choosing different branches $(e^{\pm \pi i})^{\frac{1}{\pi i \epsilon}(a+a_D)}$ for the term $(-1)^{\frac{1}{\pi i \epsilon}(a+a_D)}$ which emerges from equation (2.4) of  \cite{Basar:2015xna}. A careful matching of conventions should allow reproducing the $\im u<0$ results as well.

\section{Non-perturbative Effects in the 2d/4d Dictionary}
\label{conceptual}
Our strategy in computing the instanton partition function $\exp{\frac{1}{\epsilon_1 \epsilon_2} \cF}$ of the gauge theory in \cite{KashaniPoor:2012wb,Kashani-Poor:2013oza} was to compute the monodromy of the formal WKB solution to the null vector decoupling equation in the $\epsilon_2 \rightarrow 0$ limit as a function of $u$, and to then determine $u$ as a formal power series by imposing that this monodromy have no $\epsilon$-dependence. In this paper, using exact WKB methods, we have determined exponentially suppressed contributions $\sim \exp[ -\frac{1}{\epsilon}]$ to the relation between $u$ and the monodromy of the Floquet solutions to the differential equation. To analyze how these corrections manifest themselves in the instanton partition function, we wish to invert the relation between $u$ and the monodromy, and then invoke the definition  (\ref{identificationofu}) of $u$ in terms of $\cF$ to determine exponentially suppressed corrections to the latter.

The framework within which we will be performing these computations is that of transseries, a concept we will review in the next subsection before performing the calculation outlined above in subsection \ref{the_corrections}

\subsection{Transseries} \label{trans}
We will here consider the simplest class of transseries, which are formal power series in a finite number of generators (see for instance \cite{Costin}) -- for our purposes, these consist e.g. of $\epsilon$, $\exp[-\frac{a^{(0)}}{\epsilon}]$, and $\exp[-\frac{a_D^{(0)}}{\epsilon}]$, organized term by term with regard to the obvious relation $\gg$ (with the relative size of the last two generators immaterial). To make sense of such series, we will require the formal power series in $\epsilon$ that can be extracted at each exponential order in $\exp[-\frac{a^{(0)}}{\epsilon}]$ and $\exp[-\frac{a_D^{(0)}}{\epsilon}]$ to be Borel summable. E.g., the transseries to the left of the arrow in the expression
\be \label{map_to_fct}
e^{-\frac{a}{\epsilon}} \sum_m c_m \epsilon^m \,\longrightarrow \, e^{-\frac{a}{\epsilon}} \cS_\theta(\sum_m c_m \epsilon^m )
\ee
is to be associated to the analytic function in $\epsilon$ (within a given sector) to its right. Transseries hence allow us, at the level of formal series, to distinguish between functions with the same asymptotic power series expansion, such as $\exp[-\frac{a}{\epsilon}]$ and $0$. Furthermore, the relation $\gg$ lifts to the level of Borel transforms: an exponentially suppressed pre-factor cannot be compensated upon multiplication by the Borel transform of a formal power series, as a formal power series cannot capture the asymptotics $\sim \exp[-\frac{1}{\epsilon}]$.

When manipulating transseries, two questions naturally arise: is the result again a transseries, and can it be mapped to a function via Borel resummation (or a generalization thereof)? The former question is naturally easier to address than the latter, but also of limited usefulness: subdominant contributions are only well-defined in the context of the map (\ref{map_to_fct}), and will generically depend sensitively on the integration direction $\theta$, as we have seen. 

Transseries arise naturally in the context of exact WKB solutions. By removing the brackets $\left( \cdot \right)_\bullet$ in (\ref{finaltrace_out}), (\ref{tr2ICMS}) and (\ref{tr4ICMS}), we map our results for the trace of the monodromy matrix into transseries form. The coefficients of these transseries depend on $u$. The manipulation we wish to perform on these transseries is to solve them for $u$. As we will demonstrate in the next subsection, this is possible formally, and yields $u$ as a transseries in the generators $\epsilon$, $\exp[-\frac{\aex}{\epsilon}]$, and $\exp[-\frac{a_D(\aex)}{\epsilon}]$. A proof that this series is Borel summable is perhaps possible, combining information about the $u$-dependence of the coefficients (they are hypergeometric functions) and the growth behavior in $\epsilon$ as follows on general grounds from WKB theory, but goes beyond the confines of this work.

\subsection{$\epsilon$-instanton corrections to the instanton partition function} \label{the_corrections}
Since we are here interested in non-perturbative corrections to the instanton partition function, we will consider the large $u$ regime. Our goal is thus to solve the transseries equation\footnote{We choose the sector $(0,+)$ for convenience. The dependence of our result on the sector is a reflection of the shortcomings of transseries in the absence of a resummation result, as discussed above.} 
\be \label{a_ex_a}
e^{\frac{\aex}{\epsilon}} +e^{-\frac{\aex}{\epsilon}} = e^{\frac{a(u,\epsilon)}{\epsilon}} + e^{-\frac{a(u,\epsilon)}{\epsilon}} + e^{-\frac{1}{\epsilon} (a(u,\epsilon)+2a_D(a,\epsilon))}
\ee
underlying (\ref{finaltrace_out}) for $u$ expressed as a transseries in the generators $\epsilon$, $e^{-\frac{\aex}{\epsilon}}$, and $e^{-\frac{a_D(\aex)}{\epsilon}}$. Recall that $a(u,\epsilon)$ and $a_D(u,\epsilon)$ are defined as formal power series in (\ref{WKB_periods}). In (\ref{a_ex_a}), we have expressed $a_D$ as a function of $a$ and $\epsilon$, which is always possible at large $u$. We will proceed by first solving (\ref{a_ex_a}) to express $a$ as a function of $\aex$, and then plugging this relation into $u(a,\epsilon)$ to obtain the desired result. With the ansatz
\be \label{ansatz_for_a}
a = \aex + \epsilon \sum_{m,n=1}^\infty c_{mn} e^{-\frac{1}{\epsilon}\left(2m\, \aex + 2n\, a_D(\aex,\epsilon)\right)}  \,,
\ee
the first step amounts to solving consecutively\footnote{One possible order is to solve for all $m+n=i$ for consecutive $i$.} for the coefficients $c_{mn}$. Note that these coefficients will generically be formal power series in $\epsilon$, as they depend on derivatives of $a_D(a)$ evaluated at $\aex$. The first few terms are given by
\be \label{a_of_aex}
a= \aex - \epsilon \left( e^{-\ooe \left( 2 \aex + 2 a_D(\aex)\right) }  + e^{-\ooe \left( 4 \aex + 2 a_D(\aex)\right) }  + \frac{1}{2} (3+4 a_D'(\aex)) e^{-\ooe \left( 4 \aex + 4 a_D(\aex)\right) }   +\ldots \right) \,.
\ee
The first terms of $u(a,\epsilon)$ as a formal series in $\frac{\Lambda}{a}$ and $\frac{\epsilon_1}{\Lambda}$ can be computed based on the information provided in subsection \ref{MathieuWKB} to be
\be \label{u_of_a}
u(\asw,\epsilon) = \frac{1}{2} \asw^2 + \frac{1}{4 \asw^2} + \frac{5  }{64 \asw^6} + \ldots + \epsilon^2\left( \frac{1}{8 \asw^4} + \dots\right)  
+ \ldots \,,
\ee
where we have simplified the expression at the expense of introducing yet another $a$-variable, $\asw= \frac{\sqrt{2} a}{2 \pi i}$. The dual period $a_D$ as a function of $\asw$ is given by 
\ba
\frac{\sqrt{2}}{2 \pi i} a_D(\asw,\epsilon) & =& - \frac{1}{4 \pi i } \Big( 8 \asw \ln 2 \asw - 8 \asw + \frac{1}{\asw^3} 
+ \frac{15}{32 \asw^7} + \ldots  \nn \\
& & \qquad + \epsilon^2  \left( \frac{1}{3 \asw} +\frac{1}{ \asw^5} + \ldots \right)  
+\ldots  \Big)
   \,.
\ea
Plugging (\ref{a_of_aex}) into (\ref{u_of_a}) yields the parameter $u$ in terms of $\aex$,
\begin{eqnarray}
u&=& 
-\frac{1}{4 \pi ^2} (\aex^2 - 2  \epsilon \aex
 \exp \left(-\frac{2 \aex + 2 a_D(\aex )}{ \epsilon}\right)  
   )+O(\epsilon^2)
  \nn \\ && 
+ \,O (\text{higher non-perturbative}) \, .
\end{eqnarray}
We can restore the scale $\Lambda$ using
\begin{eqnarray}
\epsilon = \epsilon_1 / ( \sqrt{2} \Lambda ) & \mbox{and} &
\aex = \acft/( \sqrt{2} \Lambda)
\end{eqnarray} and find 
\begin{eqnarray}
u&=& 
-\frac{1}{8 \pi ^2   \Lambda^2} (\acft^2 - 2  \epsilon_1 \acft
 \exp \left(-\frac{2 \acft + 2 a_{\mbox{\tiny D,cft}} (\acft/(\sqrt{2} \Lambda) )}{ \epsilon_1}\right)  
   )+O(\epsilon_1^2)
  \nn \\ &&
+\,O (\text{higher non-perturbative}) \, .
\end{eqnarray}
Integrating $\partial_\Lambda {\cal F} = - 8 \Lambda u$ with regard to $\Lambda$ thus yields the following first 
$\epsilon_1$ non-perturbative correction to the instanton amplitude:
\begin{eqnarray}
{\cal F} &=& \frac{1}{\pi^2}
\acft^2  \log \Lambda  
+ \frac{ i}{2 \pi} \epsilon_1^2
 \exp \left(-\frac{2 \acft + 2 a_{\mbox{\tiny D,cft}} (\acft/(\sqrt{2} \Lambda) )}{ \epsilon_1}\right) + \dots \,.
\end{eqnarray}

Note that the expression (\ref{a_ex_a}) is invariant under the residual $\mathbb{Z}_2$ gauge symmetry $\aex \rightarrow -\aex$. To proceed, we choose a particular sign for $\aex$ such that the exponential
corrections in (\ref{ansatz_for_a}) are small. This choice corresponds to a gauge fixing of the residual symmetry.

\section{Conclusions and Open Problems}
\label{conclusions}

We have seen that exact WKB methods strongly suggest that the general form of correction to 
the instanton partition function $\zinst$ in an $\epsilon$-expansion be in terms of powers of $\exp[ - \frac{\aex}{\epsilon}]$ and $\exp[ - \frac{a_D(\aex)}{\epsilon}]$. We computed such corrections to $\zinst$, the result arising upon formally solving a transseries equation for one of its parameters, $u$. It remains to be shown that the formal series thus obtained for the parameter $u$ indeed corresponds to a transseries in the strong sense of subsection \ref{trans}. We checked the consistency of our computation by demonstrating independence of our result from the integration direction $\theta$ of the inverse Laplace transform. 
The specifics of the transseries form do depend on this parameter. 

The techniques used in this paper can equally well be used to compute non-perturbative corrections to superconformal gauge theories, such as $\cN=2^*$. The modularity constraints in such theories may allow for the extraction of more detailed non-perturbative information.

More broadly, we relate $\zinst$ to the null vector decoupling equation by taking the limit $\epsilon_2 \rightarrow 0$ in parameter space for which no worldsheet description of the topological string theory is yet available. Filling this gap, or adapting our methods to apply away from this limit, will be an important step towards the further physical interpretation of our results.

\section*{Acknowledgments}
We would like to thank Sujay Ashok and Benjamin Basso for useful discussions, and Reinhard Sch\"afke for correspondence.
We would like to acknowledge support from the grant ANR-13-BS05-0001.

\appendix

\section{Local Analysis of Stokes Lines} \label{local_analysis}
Consider the part of parameter space where $|u| \gg 1$. In this case, the two turning points of $Q_0$ in 
equation (\ref{q0}) lie far from the real axis, 
\be
q_{tp} \sim \pm i \log 2u + 2 \pi n  \,,
\ee
and the cosine can be well approximated by an exponential. For a turning point in the upper half-plane,
\be
\int \lambda \, dq \sim \int \sqrt{ \frac{1}{2} e^{-i (q_{tp}+x) } - u } \, dq  \sim \frac{2 i}{3u} (-ixu)^{\frac{3}{2}} \,.
\ee
The angles $\phi_x$ at which the Stokes lines emerge from the turning point are determined to first order by the condition
\be
-\frac{1}{2} \pi + \phi_x + \phi_u = \frac{2}{3} (-\theta+n \pi - \frac{1}{2} \pi + \phi_u) \,, \quad n \in \IZ \,,
\ee
with $u = |u| e^{i\phi_u}$. The lines hence emanate in the directions $\phi_x=- \frac{2}{3} \theta+ \frac{2}{3} \pi (n+ \frac{1}{4}) - \frac{1}{3} \phi_u$. Note that increasing $\theta$ results in $\phi_x$ decreasing, the Stokes lines hence rotating anti-clockwise. This implies that when decreasing an angle from a critical value, the two simple Stokes lines into which the double Stokes line splits swerve to the left, as seen from the turning point from which they emerge, whereas when increasing it, they swerve to the right. This is a property we used in subsection \ref{voros_jumps}. Let us now consider $u>0$ and $\theta=0$ for illustration purposes. In this case, the Stokes lines occur at angles $\phi_x=-\frac{1}{2} \pi, \frac{1}{6} \pi, \frac{5}{6} \pi$. Which $n$ corresponds to each of these three angles depends on the choice of branch cut, i.e. the choice of interval for the phase of the argument of the square root. Note that the pattern of Stokes lines is hence independent of the choice of branch cut, whereas the orientation of the Stokes lines does depend on this choice: for $n$ even, the Stokes line is dominant, for $n$ odd, it is recessive. To have e.g. the Stokes line in the negative imaginary direction have different parity from the other two, we can choose the argument of the radicand to lie in the interval $[ 0, 2\pi]$, corresponding to a branch cut in the positive imaginary direction. The values of $-\frac{1}{2} \pi + \phi_x $ (the phase of the argument of the square root) that fall into this interval are $\frac{1}{3}\pi, \pi, \frac{5}{3} \pi$, with $\pi$ corresponding to the negative imaginary direction. The corresponding values of $n$ are $1,2,3$, hence the distinguished Stokes line is dominant.\footnote{Note that there is a small subtlety here due to the nature of branch cuts. The branch of the square root is specified by specifying the interval within which the phase of its argument lies. If the argument is a function, we usually specify this interval for the variable of that function. But even in the simplest case of a linear function $\alpha z$, this requires specifying the phase of the factor $\alpha$ exactly, i.e. not only up to multiples of $2 \pi$. This is why we specify the branch cut not in terms of $\phi_x$, but in terms of the argument of $-ix$.}
For a turning point in the lower half-plane, we obtain
\be
\int \lambda \, dq \sim \int \sqrt{ \frac{1}{2} e^{i (q_{tp}+x) } - u } \,dq \sim -\frac{2i}{3u} (ixu)^{\frac{3}{2}} \,,
\ee
hence with the same choice of branch cut, $\phi_x= -\frac{1}{6}\pi, \frac{1}{2} \pi, \frac{7}{6} \pi$, corresponding to $n=0,1,2$. The Stokes line pointing in the positive imaginary direction is hence recessive.

\section{Numerical Results} \label{num_res}
In this appendix, we will compare results for the Mathieu characteristic exponent obtained numerically with those attainable by evaluation of the formulae determined in subsection \ref{bpt}.

\subsection{Numerical WKB perturbation theory}
\label{WKBpertnum}
In a first numerical experiment, let us illustrate the efficacy of WKB perturbation theory.
Consider the following values of $u$ and $\epsilon$:
\be
\epsilon = \frac{1}{10} \,, \quad u = 2 + 2 i \, .
\ee
The value of $\epsilon$ is small compared to $1$ and the value of $u$ lies outside the curve of marginal stability. The values of $a$ and $2a_D-a$ at leading order in perturbation theory are\footnote{In this appendix, the symbol for a formal power series with superscript ${}^{(n)}$ will indicate that power series numerically evaluated up to the $n$-th order term.}
\be
a^{(0)}/\epsilon \sim 41-97 i\,, \quad (2a_D-a)^{(0)} /\epsilon\sim -90 + 19 i \,.
\ee
Non-perturbative effects are hence heavily suppressed, and the Mathieu characteristic exponent should be well approximated by the $a$-period of the differential $\lambda$ and its higher order corrections in $\epsilon$.
Indeed, the Mathematica routine for the numerical approximation to the Mathieu exponent gives the value \footnote{We record the Mathematica value times $-i \pi$ to render normalizations uniform.} (up to 20 digits)\footnote{Here and in the following, we do not round the numerical values.}
\begin{eqnarray}
\mu &=& 41.209556582920410400-97.333703725326654896 i \, .
\end{eqnarray}
The successive approximations using the WKB formulae (\ref{WKBpert}) in $\epsilon$-perturbation theory are
\begin{eqnarray}
a^{(0)} /\epsilon&=& 41.20893-97.33334 i  \,,
\nonumber \\
a^{(2)} /\epsilon&=& 41.20955667-97.33370309 i  \,,
\nonumber \\
a^{(4)} /\epsilon&=& 41.2095565837-97.33370372575 i  \,,
\nonumber \\
a^{(6)} /\epsilon&=& 41.2095565829175-97.3337037253288 i   \,,
\nonumber \\
a^{(8)} /\epsilon&=& 41.209556582920406-97.33370372532663 i \, .
\end{eqnarray}
Each iteration enhances the accuracy by approximately two digits, which is what one would naively expect of a perturbation series in $\epsilon^2 =10^{-2}$.
These numerical results demonstrate the usefulness of WKB perturbation
theory, even in its asymptotic form.

\subsection{Numerics and the Stokes corrected formula}\label{nands}

In this subsection, we will consider examples for which the non-perturbative effects computed in subsection \ref{bpt} become numerically significant. 
The evaluation of the formulae derived there will be sector dependent, as we will approximate Voros multipliers by the first terms in their asymptotic expansion. We will work inside the curve of marginal stability. Of the two sectors that occur here, we will observe that one yields the better approximation to the characteristic exponent, indicating that the Voros multiplier in this particular sector is better approximated by its asymptotic series.

Let us consider the values
\be
\epsilon = \frac{1}{5}  \,, \quad u =\frac{1}{3} e^{- i \frac{\pi}{10}}\, .
\ee
We shall compare the value for the trace of the monodromy matrix determined by Mathematica \cite{MaMa} to those which follow from the numerical evaluation of the perturbative WKB result and the exact WKB result in the sectors $(++-)$ and $(+--)$, to $0^{th}$ and $4^{th}$ order in perturbation theory.
\begin{eqnarray}
\tr M_A^{\mbox{\tiny Mathematica}} &=&
3078.40577   - 11972.57629 i\,,
\nonumber \\
\left(2 \cosh \frac{a}{\epsilon}\right)^{(0)} &=& 3313.34282   - 11850.61433 i\,,
\nonumber \\
\left(2 \cosh \frac{a}{\epsilon} + e^{\frac{1}{\epsilon}(2 a_D-a)} \right)^{(0)} &=& 2810.06097    -12265.24243 i\,,
\nonumber \\
\left(2 \cosh \frac{2a_D-a}{\epsilon} + e^{\frac{1}{\epsilon}a} \right)^{(0)} &=& 2810.05976   -12265.24153 i\,,
\nonumber \\
\left(2 \cosh \frac{a}{\epsilon}\right)^{(4)} &=& 3582.79033    - 11581.22299 i \,,
\nonumber \\
\left(2 \cosh \frac{a}{\epsilon} + e^{\frac{1}{\epsilon}(2 a_D-a)} \right)^{(4)}&=& 3078.40503    -11972.57314 i  \,,\nn \\
\left(2 \cosh \frac{2a_D-a}{\epsilon} + e^{\frac{1}{\epsilon}a} \right)^{(4)} &=& 3078.40377    -11972.57226 i \, .
\end{eqnarray}
The formulae for the two sectors $(++-)$ and $(+--)$ yield numerically close results, as we are approximating
\ba
\left(2 \cosh \frac{a}{\epsilon} + e^{\frac{1}{\epsilon}(2 a_D-a)} \right)_{(++-)} &\sim& e^{\frac{1}{\epsilon} a^{(4)}} + e^{\frac{1}{\epsilon}(2 a_D-a)^{(4)}} + e^{-\frac{1}{\epsilon} a^{(4)} }  \,, \label{approx_ppm}\\ 
\left(2 \cosh \frac{2a_D-a}{\epsilon} + e^{\frac{1}{\epsilon}a} \right)_{(+--)} &\sim& e^{\frac{1}{\epsilon} a^{(4)}} + e^{\frac{1}{\epsilon}(2 a_D-a)^{(4)}} + e^{-\frac{1}{\epsilon}( 2 a_D-a)^{(4)}} \,, \label{approx_pmm}
\ea
with 
\be
a^{(4)} /\epsilon=  9.40284+ 17.5788 i \,, \quad  (2a_D-a)^{(4)}/\epsilon = 6.45897-15.0481i\,.
\ee
To be able to better discriminate numerically between the two equations (\ref{approx_ppm}) and (\ref{approx_pmm}), we shall next choose a value of $u$ at which $|2a_D-a|$ is small, and then choose the phase of $\epsilon$ to eliminate the real part of $a$.

To this end, let us choose 
\be 
u= -\frac{4}{5}i \,.
\ee
Then 
\be
a^{(4)}=4.23320+4.3172i = 6.04636 \,e^{i \pi (0.25312)} \,, \quad  (2a_D-a)^{(4)}/\epsilon =0.04966-0.28854i \,.
\ee
Choosing $\epsilon$ to rotate away the phase of $a$,
\be
\epsilon = \frac{1}{10} \, e^{i \pi (\frac{1}{4} + \frac{1}{2})} \,,
\ee
we obtain
\begin{eqnarray}
\tr M_A^{\mbox{\tiny Mathematica}}&=& 9.84414 \,,
\nonumber \\
\left(2 \cosh \frac{a}{\epsilon}\right)^{(0)} &=& -1.46994 \,,
\nonumber \\
\left(2 \cosh \frac{a}{\epsilon} + e^{\frac{1}{\epsilon}(2 a_D-a)} \right)^{(0)}  &=& 10.03070  \,,
\nonumber \\
\left(2 \cosh \frac{2a_D-a}{\epsilon} + e^{\frac{1}{\epsilon}a} \right)^{(0)}  &=& 10.85260 +0.67809 i \,,
\nonumber \\
\left(2 \cosh \frac{a}{\epsilon}\right)^{(4)}&=& -1.46174 \,,
\nonumber \\
\left(2 \cosh \frac{a}{\epsilon} + e^{\frac{1}{\epsilon}(2 a_D-a)} \right)^{(4)} &=& 9.84414 \,, \nn \\
\left(2 \cosh \frac{2a_D-a}{\epsilon} + e^{\frac{1}{\epsilon}a} \right)^{(4)} &=& 10.66350+0.68251 i\, .
\end{eqnarray}
For this example, the approximate evaluation of our exact WKB formulae in the sectors $(+--)$ and $(++-)$ yields appreciably different values, and one sector, $(++-)$, yields the better approximation. This indicates that the Voros multipliers are better approximated by their asymptotic expansion in this sector.

\section{Periodicity, Determinant, and Numerics}  \label{mama_hm}
In this section, we review the determinant formula for the exact periodicity, an efficient
way to evaluate the determinant, and how the resulting numerics may improve on a built-in
Mathematica \cite{MaMa} evaluation of the Mathieu characteristic exponent.
\subsection{Hill's method}
\label{Hill}
Hill's method yields an exact formula for the Mathieu characteristic
exponent $\nu$ in terms of the parameters in the Mathieu equation (see e.g. \cite{Strang} and references therein).

The derivation of this formula proceeds as follows. The power series
\be
e^{i \nu q/2} p(z) = e^{i \nu q/2} \sum_{r=-\infty}^{+\infty} c_{2r} e^{i r q}
\ee
provides a formal solution to Mathieu's equation (\ref{math_diff}) if its coefficients satisfy
\begin{eqnarray}
2 c_{2r-2} + ( \epsilon^2 ( 2r +\nu)^2 - 4u ) c_{2r} + 2  c_{2r+2} &=& 0 \, .
\end{eqnarray}
By dividing all coefficients in this recursion relation by the coefficient of the second term,
the determinant $\Delta(\nu)$ of the matrix underlying this linear set of equations for the coefficients $c_r$ becomes convergent \cite{Strang}. For a non-trivial solution to this infinite
set of equations to exist, $ \Delta(\nu)$ must vanish. The determinant is invariant under $\nu \rightarrow 2 n \pm \nu$ (with $n$ integer),
and is therefore even and periodic in $\nu$ with period $2$. It has simple poles
at $(2r +\nu)^2 - 4u \epsilon^{-2} =0$, and it tends to $1$ as $\nu \rightarrow \infty$. This behavior determines, by Liouville's theorem, the form of $\Delta(\nu)$ up to a constant \cite{Strang} 
\be
1-\Delta(\nu)  = \frac{(\Delta(0)-1) \sin^2(\pi \sqrt{u \epsilon^{-2}})}{\sin^2 \frac{\pi \nu}{2} - \sin^2( \pi \sqrt{u \epsilon^{-2}})} \,.
\ee
Imposing $\Delta(\nu)=0$ then gives rise to the following constraint equation on $\nu$:
\be
\sin^2 \frac{\pi \nu}{2} = \Delta(0) \sin^2 {\pi \sqrt{u\epsilon^{-2}}}\, ,
\ee
or equivalently
\be
\cos \pi \nu = 1 - 2 \Delta(0) \sin^2 {\pi \sqrt{u \epsilon^{-2}}} \, .
\ee

\subsection{Linear Recursion}\label{tri}
The determinant $\Delta(0)$ can be computed efficiently as it satisfies
\begin{eqnarray}
\Delta(0) &=& \det A\,,
\end{eqnarray}
where $A$ is a tri-diagonal matrix \cite{Strang}. We denote the determinant of the submatrix of size $(2i+1) \times (2i+1)$ acting on $(c_{-2i}, \ldots, c_{-2},c_0,c_{2}, \dots, c_{2i})^T$ 
as $\Delta_i$. In the limit $i\rightarrow \infty$, we recover $\Delta$.
The determinants $\Delta_i$ 
satisfy a linear recursion relation (proved by computing minors)
\begin{eqnarray}
\Delta_i &=& (1-\alpha_i) \Delta_{i-1}  - \alpha_i (1-\alpha_i) \Delta_{i-2} + \alpha_i \alpha_{i-1}^2 \Delta_{i-3} \, ,
\end{eqnarray}
with
\begin{eqnarray}
\alpha_n &=& \frac{1 }{4 ( n^2 \epsilon^{2} -  u ) ( (n-1)^2  \epsilon^{2} -  u  )} \, ,
\end{eqnarray}
and initial terms
\begin{eqnarray}
\Delta_0 &=& 1 \,,
\nonumber \\
\Delta_1 &=& \det 
\left( 
\begin{array} {ccc}
	1 & \xi_2 & 0 \\
	\xi_0 & 1 & \xi_0 \\
     0 & \xi_{-2} & 1	
\end{array} 
\right) \,,
\nonumber \\
\Delta_2 &=&  \det 
\left( 
\begin{array} {ccccc}
 1 & \xi_4 &  0 &  0 & 0 \\
	\xi_2  & 1 & \xi_2 & 0 & 0 \\
	0 & \xi_0 & 1 & \xi_0 & 0 \\
     0 & 0 & \xi_{-2} & 1& \xi_{-2} \\
     0 & 0 &  0 & \xi_{-4} &  1	
\end{array}
\right) \, ,
\end{eqnarray}
where
\begin{eqnarray}
\xi_n &=& \frac{2}{n^2 \epsilon^2- 4 u } \, .
\end{eqnarray}
Our implementation of this linear recursion in Mathematica demands a computation time that grows linearly in the size of the matrix $A$. 
The reduced cost is due to the fact that $A$ is a tridiagonal matrix.

\subsection{The Determinant, Mathematica, and WKB}
Mathematica provides a convenient numerical algorithm for 
finding the Mathieu characteristic exponent for any complex parameters of the Mathieu equation. However, the algorithm is a black box. In our numerical experiments, we have found that the algorithm must be used with caution, in particular at small values of $|\epsilon|$, as the following example demonstrates. We will compute
the characteristic exponent in a particular case using three methods, namely Mathematica, WKB perturbation theory, and the numerical algorithm based on Hill's method describe in subsections \ref{Hill} and \ref{tri}.

We consider the values
\be
\epsilon = 10^{-2} \,, \quad  u = 6 e^{i \frac{\pi}{8}} \, ,
\ee
and have Mathematica compute the parameters
\be
 a_M = 4 u / \epsilon^2 \,,\quad q_M = 2/ \epsilon^2  
\ee
and the characteristic exponent
\be 
   \mbox{N}[-i \pi \mbox{MathieuCharacteristicExponent}[a_M, q_M], 20]] \,,
\ee
yielding the result
\be \label{mathematica_res}
301.3467972919577793-1507.3089399330755493  i \, .
\ee
The WKB approximation to fourth order for the a-period yields
\be  \label{WKB_res}
301.7573905535926129-1507.2622454539553016  i \, .
\ee
At small $\epsilon$, we expect to obtain a better approximation to the characteristic exponent than the comparison of (\ref{mathematica_res}) and (\ref{WKB_res}) would suggest, given that $a/\epsilon \sim 302-1507i$ and $2a_D-a \sim -2020+770i$ imply highly suppressed non-perturbative corrections.

To check the Mathematica result, we programmed a numerical algorithm based on Hill's method.
For a matrix of size $2\times 10^5$, we found the result
\begin{eqnarray}
301.75739055567594974+ 0.70222826914545431  i \, .
\end{eqnarray}
This agrees with WKB perturbation theory to high order 
-- the difference in the imaginary part is an integer multiple of $2 \pi$, which corresponds to
the ambiguity in the Mathieu characteristic exponent. Thus, for small $\epsilon$, the  perturbative WKB approximation is more reliable than the (unspecified) Mathematica algorithm. It agrees with an alternative numerical algorithm which has linear cost in the size of the (sparse) matrix used in the evaluation of the determinant.

\bibliography{exact}
\bibliographystyle{utcaps}

\end{document}